\documentclass[
 reprint,
 amsmath,amssymb,
 aps,
]{revtex4-1}

\usepackage{color}
\usepackage{mathtools}
\usepackage{amsmath,amssymb}
\usepackage{graphicx}
\usepackage{dcolumn}
\usepackage{bm}
\usepackage{float}
\usepackage{epstopdf}
\usepackage{textcomp}

\def\be{\begin{equation}}
\def\ee{\end{equation}}
\def\bea{\begin{eqnarray}}
\def\eea{\end{eqnarray}}
\def\bsplit{\begin{split}}
\def\esplit{\end{split}}

\def\nn{\nonumber}
\def\f{\frac}

\def\l{\left(}
\def\r{\right)}
\def\a{\alpha}
\def\d{\nabla}
\def\bs{\boldsymbol}
\def\ddx{\partial_x}
\def\ddxx{\partial^2_x}

\def\m{\Delta\mu}
\def\e{\epsilon}
\def\[{\left[}
\def\]{\right]}

\newcommand{\ncbs}{\affiliation{Simons Centre for the Study of Living Machines, National Centre for Biological Sciences (TIFR), Bangalore 560065, India}}
\newcommand{\rri}{\affiliation{Raman Research Institute, Bangalore 560080, India}}
\newcommand{\ibdml}{\affiliation{IBDM, UMR7288, CNRS-Aix Marseille Universite, Campus de Luminy case 907, 13009 Marseille, France}}

\begin{document}

\preprint{APS/123-QED}

\title{Actomyosin pulsation and symmetry breaking flows in a confined active elastomer subject to affine and nonaffine deformations}

\author{Deb Sankar Banerjee}
\rri

\author{Akankshi Munjal, Thomas Lecuit}
\ibdml

\author{Madan Rao}%
\rri
\ncbs

\date{}
           
\begin{abstract}
Tissue remodelling in diverse developmental contexts require cell shape changes that have been associated with pulsation and flow of the actomyosin cytoskeleton. Here we describe 
the dynamics of the actomyosin cytoskeleton 
as a confined active elastomer embedded in the cytosol and subject to 
turnover of its components. 
Under {\it affine} deformations (homogeneous deformation over a spatially coarse-grained scale), the active elastomer 
 exhibits 
  spontaneous oscillations, propagating waves, contractile collapse and spatiotemporal chaos. The collective nonlinear dynamics shows nucleation, growth and coalescence of  actomyosin-dense regions
  which, beyond a threshold, spontaneously move as a 
  {\it spatially localized traveling front}
 towards one of the boundaries. However, large 
myosin-induced contractile stresses, can lead to {\it nonaffine} deformations due to actin turnover.
 This 
results in a {\it transient actin network}, that naturally accommodates  {\it intranetwork flows} of the 
actomyosin dense regions as a consequence of filament unbinding and rebinding.
 Our work suggests that the driving force for the spontaneous movement comes from the actomyosin-dense region itself and not the cell boundary.
 We verify the many predictions of our study in
{\it Drosophila} embryonic epithelial cells undergoing neighbour exchange during a collective process of tissue extension called germband extension.

\end{abstract}

\pacs{Valid PACS appear here}

\maketitle

\section{Introduction}

\noindent
Tissue remodelling in diverse developmental contexts, such as apical constriction 
in {\it Drosophila} \cite{Martin,solonbrunner,meghana,blanchard}, or {\it C. elegans} \cite{rohjohnson},
cell intercalation
during {\it Drosophila} germ band extension \cite{Lecuit2010,Lecuit2013,akankshi,zallen}, and xenopus extension and convergence \cite{xenopus}
 have been
observed to be associated with pulsation and flows of the medial actomyosin cytoskeleton. For instance,
tissue extension in the {\it Drosophila} embryo proceeds by the intercalation of cells, a so called T$_1$-process, which starts with
the step-wise shrinkage of  a subset of cell junctions called `vertical junctions' (aligned with the dorsal-ventral axis of the embryo) driven by cortical Myosin-II.
 It was recently shown that each step of junctional shrinkage is driven by a medial-apical actomyosin pulse that flows towards this vertical junction
 \cite{Lecuit2010,Lecuit2008,akankshi}. 
In this paper, we describe a general theory of the dynamics of spontaneous actomyosin pulsation and symmetry breaking flows, that should be
applicable not only in the context of germ-band elongation, but also during other morphogenetic events.
In addition,
using germ band cells in the {\it Drosophila} embryo as a model system, we provide experimental justification for the assumptions underlying our theoretical framework and verify many of its key predictions.

\begin{figure*}
\centering
 \includegraphics[width=120mm]{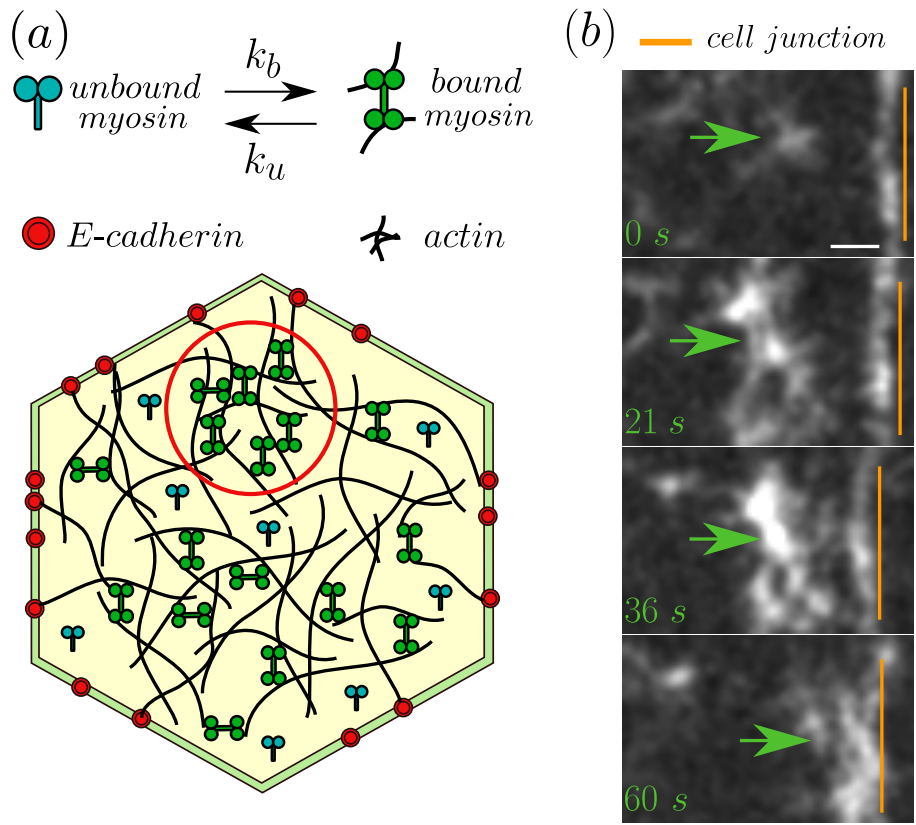}
    \caption{(a) Schematic showing the medial actomyosin cytoskeletal meshwork within a cell belonging to the tissue. The actin filaments are attached to the cell junctions via E-cadherin (red dots). 
      Myosin minifilaments bind (unbind) with rates $k_b (k_u)$ and apply contractile stresses on the actin filament meshwork, the red circle demarcates a region of
      higher mesh compression.
  Both the actin filaments and the myosin minifilaments undergo turnover. (b) Time snapshots  showing the nucleation, growth and coalescence 
  of an
  actomyosin-dense cluster and its subsequent flow towards the cell junction, in a germ band cell of the drosophila wing imaginal disc. Myosin has been labeled with RLC-mCherry. Scale bar $1 \mu$m.
	    }
	    \label{fig:schema}
\end{figure*}

The medial actin mesh modelled as an active elastomer embedded in a solvent is subject to active contractile stresses arising from the binding of myosin
mini-filaments (Fig.\,\ref{fig:schema}a) \cite{Lecuit2013,Martin_jcb2010,rohjohnson,Wu_NCB_2014}. The actin mesh is connected to E-cadherin adhesion molecules  at the  cell junctions via molecular linkers such as $\alpha$-catenin 
an actin binding protein \cite{cavey,yonemura,rohjohnson,buckley} and $\beta$-catenin which binds $\alpha$-catenin and E-cadherin. 
Viewing the high resolution time lapse images (Fig.\,\ref{fig:schema}b) or movies ({\it Supplementary Movies}) showing nucleation, growth, coalescence and flow of clusters of labeled myosin towards the cell junction in a {\it Drosophila} germ band cell, should
immediately convince one of the utility of adopting a hydrodynamic approach. 
The hydrodynamic
equations for this active elastomer are derived from symmetry arguments and include minimal phenomenological inputs
(Section II). We describe the hydrodynamic modes in the overdamped limit \cite{Marchetti2011a,Marchetti2011b} and obtain phase diagrams by numerically solving the hydrodynamic equations (Section III).
We find that the active elastomer exhibits spontaneous oscillations, contractile instabilities and coarsening of clusters enriched in actomyosin.
The coarsening often leads to stable actomyosin-dense clusters which, beyond a threshold,  
acquire a polarity. This results in a spontaneous movement as a spatially localised traveling front.
Such localised traveling front solutions appear in other
excitable systems such as the Fitzhugh-Nagumo model \cite{keener}, to which our model bears a close resemblance.
On the other hand, large 
myosin-induced contractile stresses, can lead to {\it nonaffine} deformations due to actin turnover (Section IV).
 This 
results in a {\it transient actin network}, that exhibits  {\it intranetwork flow} of the 
actomyosin dense regions as a consequence of filament unbinding and rebinding.
 Interestingly, the theory predicts that
the driving force for spontaneous flow comes from the actomyosin-dense region itself and not the cell boundary - 
we provide robust experimental verification 
of this. 
While our hydrodynamic analysis is designed to be very general, the affine description should be
seen as being closely related to the spring model of \cite{SolonSalbreux}. Our general perspective reveals several 
significantly new aspects and provides a fresh conceptual understanding of this ubiquitous phenomenon.

\section{Model and Equations of Motion}
The hydrodynamic variables of the confined active elastomer are the actin mesh affine displacement field ${\bf u}$, the actin mesh density $\rho$, the density of the bound myosin 
minifilaments $\rho_b$,
 the junctional E-cadherin density $\rho_{c}$ and the hydrodynamic velocity $v$ (Fig.\,\ref{fig:schema}a).

In the absence of flow and activity, the system relaxes to equilibrium governed by a free-energy functional $F[{\bs u}, \rho, \rho_b, \rho_{c}]
=\int d{\bf r} f_B$, where  $f_B$ is the the bulk elastic free energy density
of the elastomer written in terms of the linearized
strain tensor  $\boldsymbol \epsilon = \frac{1}{2} (\boldsymbol \nabla \boldsymbol u + (\boldsymbol \nabla \boldsymbol u)^{T})$
({\it Methods}).  In principle, we need to include a boundary contribution
 reflecting the 
soft anchoring arising from the actin-cadherin linkage; for the present we simply include $\rho_c$'s  contribution in the
elastic moduli 
and
hard boundary conditions on the dynamical equations.

In the presence of active processes,
  the overdamped Rouse dynamics
  are described by,
\begin{equation}
\Gamma \dot{\boldsymbol u}  =  \boldsymbol \nabla \cdot (\boldsymbol \sigma^e + \boldsymbol \sigma^a 
                                                                           +\boldsymbol \sigma^{d})
\label{eq:mesh}
\end{equation}
\begin{equation}
\dot{\rho_b} + \boldsymbol \nabla \cdot (\rho_b \dot{\boldsymbol u})=  D {\nabla}^2 \rho_b +  {\cal S}_m
\label{eq:myosin}
\end{equation}
\begin{equation}
\dot{\rho} + \boldsymbol \nabla \cdot (\rho \dot{\boldsymbol u}) =  {\cal M} \nabla^2 \frac{\delta F}{\delta \rho} + {\cal S}_a
\label{eq:actin}
\end{equation}
Eq.\,(\ref{eq:mesh}) is a balance between frictional force experienced by the mesh (with coefficient $\Gamma$) and the net force acting on the mesh, written in terms of
 the total stress  $\boldsymbol \sigma\equiv \boldsymbol \sigma^e + \boldsymbol \sigma^a 
                                                                           +\boldsymbol \sigma^{d}$,

                                                                            a sum of the
 elastic stress, $\boldsymbol \sigma^e = \frac{\delta F}{\delta \boldsymbol \epsilon}$, dissipative stress due to the viscosity ($\eta$) of the elastomer network, 
 $\boldsymbol\sigma^{d}=\eta \nabla{\dot{\bs u}}$, and the active stress

 \be
{\bs \sigma}^a = - \zeta(\rho, \rho_b, \bs \epsilon) \Delta \mu \,\bs I
=  - \f{\zeta_1 \rho_b}{1+\zeta_2 \rho_b}  \chi(\rho) \Delta \mu \,\bs I
\label{eq:activephen}
\ee
where $\Delta \mu$ is the difference in chemical potential during ATP hydrolysis, $\chi(\rho)$ is some smooth function of the mesh density, $\bs I$ the identity matrix
and the active stress parameters $\zeta_1<0$ (contractility!) \cite{rmp} and $\zeta_2 > 0$.
 Eq.\,(\ref{eq:myosin}) describes the dynamics of bound myosin filament density from advection by the filament velocity ${\dot {\bf u}}$ and 
turnover ${\cal S}_m=-k_u(\bs \epsilon) \rho_b + k_b \rho$ by (un)binding. We allow for a possible strain-induced unbinding
with a Hill-form,
$k_u(\bs \epsilon)= k_{u0} e^{{\bs \alpha}\cdot{\bs \epsilon}}$  \cite{kovacs,Eaton_devCell}. Similarly,
Eq.\,(\ref{eq:actin}) describes the dynamics of the actin mesh density including advection, permeation and actin turnover, taken to be  ${\cal S}_a = \l k_{+} - k_{-}\r \rho$.

In our analysis we have
ignored the hydrodynamics of the fluid, this is suggested by experiments that show the movement of the actin mesh relative to the fluid does not advect particles suspended
in the fluid unless they are bound to the actomyosin mesh \cite{Lecuit2010,akankshi}. 
For completeness however, we display the full hydrodynamic equations  in {\it Supplementary Information (SI)}.

It is important to note that in writing Eqs.\,(\ref{eq:mesh}-\ref{eq:actin}), we have tacitly taken the mesh deformations to be affine - i.e., 
where the deformations are homogeneous over a spatially coarse-grained scale. This in particular implies that the deformation does not alter
the local coordination number of the mesh.
We will return to this important point 
later.
Even so there is a clear separation of time scales, which in increasing order correspond to 
the advection time scale, myosin turnover time and actin turnover time.
We will first consider the case when the actin mesh density is ``fast'' compared to $P_e\equiv L_v V/{\cal M}$ ($L_v$ is the characteristic scale on which the 
mesh velocity, of typical magnitude $V$, varies) and the actin turnover time is long; in this limit, the mesh density is slaved to the local 
compressive strain,
thus with $\rho = \rho_0 + \delta \rho$, we arrive at $\delta \rho \propto - \epsilon_{ii}$ (see {\it Methods}).

\section{Results}
 The conceptual features of the pulsation and flow are captured by a simple scalar version with one-elastic constant, $\boldsymbol\sigma^e = B \bs \epsilon$. 
It helps to make the equations dimensionless by choosing length and time in units of $l=\sqrt{\eta/\Gamma}$ and $k_b^{-1}$, respectively; in these
units $u/l \to u$, $\rho_b/\rho_{b0} \to \rho_b$, $B/\Gamma k_b l^2 \to B$, $\zeta_1\m\rho_{b0}/k_b \Gamma l^2\to \zeta_1 \m$, $D/k_b l^2 \to D$
and $k_{u0}/k_b \to k$ are dimensionless. See {\it Methods} for parameter values.

\subsection{Linear Analysis}
We first do a linear stability analysis about the unstrained, homogeneous steady state. To this order, the active stress 
$\sigma^a=-\zeta_1 \Delta \mu (1+ \zeta^{\prime} \rho) \rho_b$, with $\zeta_1<0$  for contractility and $\zeta^{\prime}>0$.
By going over to a plane wave basis $e^{\pm i {\bf q}\cdot {\bf x}}$ (where, $2\pi/L \leq \vert {\bf q}\vert \leq 2\pi/l$, $L$ and $l$ being the system and mesh size, respectively) and computing the two roots of the dispersion (detailed calculations and phase diagram appear in the {\it SI}),

we deduce the following : (i) for low values of active stress compared to the elastic stiffness, $ -\zeta_1 \m < (B + D(1+\sqrt{k/D})^2)/2$, the elastic mesh is stable with dispersion $\omega \sim q^2$, (ii) as the contractile strength 
exceeds $(B + D(1+\sqrt{k/D})^2/2$, the elastomer undergoes unstable oscillations with an amplitude that  increases  exponentially with time, (iii) at a
threshold boundary, the elastomer supports a traveling wave solution with a speed $v_c^* =\sqrt[4]{\l kDB^2/4\r}$, (iv) beyond an active
stress $-\zeta_1 \Delta \mu \sim B$, the elastomer contracts indefinitely. 
We find the qualitative features of the transitions remain unaltered when the strain-dependent unbinding parameter $\alpha$ is varied, as long as $\alpha \leq \alpha_{max}(B, \zeta_1\m)$.

In spite of the simplifying nature of the linear analysis, it captures some gross features and makes useful predictions : 
(a) the oscillatory behaviour or pulsation requires advection and myosin turnover \cite{akankshi,Vasquez_2014}, (b) the contractile instability is promoted by lowering elastic 
stiffness (by reducing levels of $\beta$-catenin/cadherin) \cite{Lecuit2013,rohjohnson,Martin_jcb2010} or reducing myosin unbinding rates \cite{akankshi,Vasquez_2014,Kasza_KE_PNAS14}. The linear analysis however fails to capture the phenomenology of flowing configurations
of actomyosin.

\subsection{Effect of nonlinearities : stable pulsation and flows}
It is possible that the oscillatory  instability seen in the linear analysis would be tempered by nonlinearities in the dynamical equations - to lowest order these arise from the
 strain-dependent unbinding $k_u(\epsilon)$, advection, and active stress, which together 
 constitute an excitatory-inhibitory 
 system of equations leading to sustained spontaneous oscillations. To see this, we work in 1-dim, 
 and after fourier transforming Eqs.\,(\ref{eq:mesh}),(\ref{eq:myosin}) in a finite domain $[0,L]$ with Neumann boundary conditions, retain only the
 smallest wavenumber, a 
 1-mode Galerkin truncation, which accomodates the strain-dependent unbinding nonlinearity.
 The resulting coupled ODEs upon re-scaling 
 describe a generalized Van 
der Pol oscillator \cite{strogatz} with linear damping and cubic nonlinearities. This admits a limit cycle through a supercritical-Hopf bifurcation 
for $-\zeta_1\m \geq \f{B}{2}+\f{1}{\pi^2}$ - a signature of the appearance of sustained spontaneous oscillations. At the onset of bifurcation, we 
use  a fluctuation analysis to obtain the
time period of oscillation $T = 2\pi\l \eta/k_{u0}(B+\zeta_1\m\rho_{b0})\r^{1/2}$.
To see the effects of the {\it advective nonlinearity}, we need to extended the above mode-truncation analysis to 2-modes. 
 The resulting 3-dimensional 
dynamical system exhibits, in addition to limit cycles, {\it temporal chaos} as seen by the algebraic decay of the power-spectrum, positive Lyapunov
exponent  and denseness of the 
Poincare section  (details will appear in \cite{debsankar}). However, again including nonlinear effects using low order mode-truncation, fails to capture the phenomenology of flowing configurations
of actomyosin.

Note that wave-like dispersion relations obtained from this truncated model are not a response to an external perturbation, but are
{\it self-generated}. This {\it cell-autonomous} behaviour is consistent with observations of pulsatile dynamics in medial actomyosin  \cite{Lecuit2008,Lecuit2010}.

   \begin{figure*}
   \centering
   \includegraphics[width=120mm]{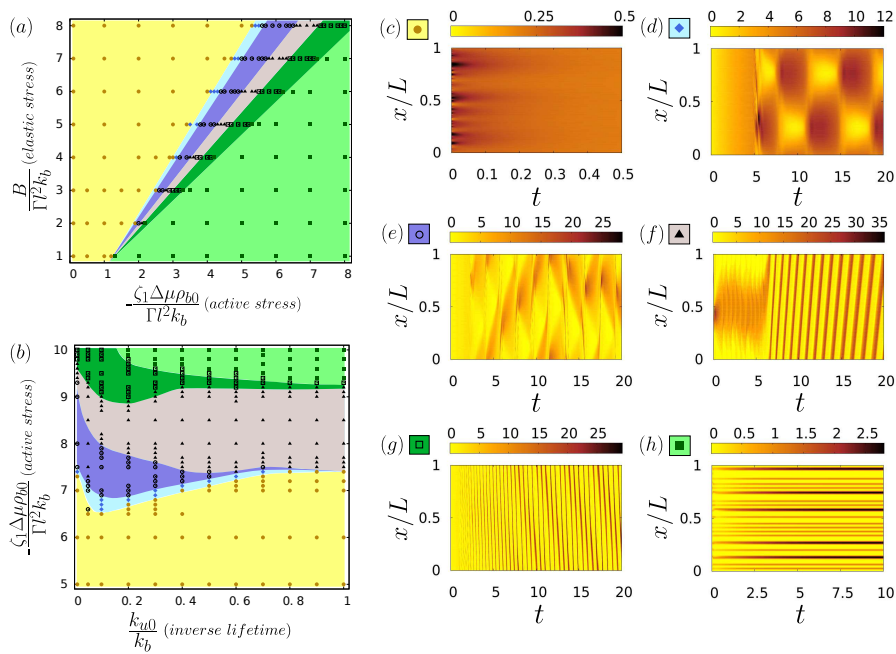}
   \caption{Phase diagram obtained from numerical solutions of Eqs.\,(\ref{eq:finalu}-\ref{eq:finalm}) : (a) Effective elastic stress density vs. contractile stress density,
   with $k=0.10$.
  (b)  Effective contractile stress density vs. inverse-lifetime, with $B=5.0$. The phases are (i) Stable (yellow), (ii) spontaneous Oscillatory (blue), (iii) spontaneous Moving (grey) and (iv) contractile Collapse (light-green). The corresponding kymographs 
  of the bound myosin 
  density (indicated by the
  colour and symbol on the upper right corner) is shown in (c), (d), (f) and (h), respectively. The regions marked violet and dark-green  are the coexistence phases - 
  the oscillatory-moving coexistence (open circle) and the collapse-moving  coexistence (open square), with the corresponding kymographs shown in 
  (e) and (g), respectively. Apart from the new phases, the topology of the phase diagrams are roughly similar to the linear stability diagram (Fig.\,S1),
    except for the upturn of the phase boundaries towards larger active stress in (b), which arises from the non-linear strain-dependent unbinding.
Symbols are points at which numerical solutions have been obtained. Rest of the parameters are, $\alpha=3.0$, $c=0.1$, $\chi(\rho_0)\zeta_1=-0.5$ and $\zeta_2=0.1$
({\it Methods}).
	    } 
	    \label{fig:numericalPD}
   \end{figure*}
   We now study the effect of the nonlinearity arising from the active stress. 
 This gives  rise to a new set of solutions, namely the {\it spatially localized traveling front} solutions.
To see this, we perform a numerical analysis of the full nonlinear equations in 1-dim.
We first 
Taylor expand $\chi(\rho)$ in Eq.\,(\ref{eq:activephen}) about $\rho_0$, the mesh density in the unstrained configuration, and recast the the active stress as
\be
\sigma^a =   \f{-\zeta_1 \Delta \mu \rho_b}{1+\zeta_2 \rho_b}  \l  \chi(\rho_0) - \f {C}{A} \chi^{\prime}(\rho_0) \epsilon + \l\f{C}{A}\r^2 \chi^{\prime\prime}(\rho_0) \epsilon^2 
+ \ldots \r
\label{eq:sigmaa}
\ee 
Separating out the terms dependent on $\epsilon$ and 
 only on $\rho_b$, and combining the former with the elastic stress $\sigma^e=B\epsilon$ in Eq.\,\ref{eq:mesh}, leads to an 
  effective ``elastic free-energy'',
\be
\Phi(\epsilon) = \frac{1}{2}K_2(\rho_b, \rho_0)\epsilon^2 + \frac{1}{3}K_3(\rho_b, \rho_0) \epsilon^3 + \frac{1}{4}K_4(\rho_0)\epsilon^4
\ee
where the $K$'s are density dependent coefficients and we have included a quartic term with $K_4>0$ to ensure that the  local compressive strain does not grow without bound, as a consequence of  steric hinderance,  filament rigidity or crosslinking myosin. The $\Phi(\epsilon)$ that emerges as a consequence of activity, has 3 new features : (i) for weak active stress, the minima at $\epsilon=0$ gets shallower, indicating that the elastic stiffness $B$ decreases, (ii) as we increase the active stress, there appears another minimum at $\epsilon=\epsilon_0$ (iii) for large active stresses, the $\epsilon=0$ state can be unstable, with the effective $B<0$ (Fig.\,S3a). 
The final 1-dim equations of motion are given by,
\bea
\label{eq:finalu}
\Gamma \dot{u}  &=& \ddx \Phi^{\prime}(\epsilon) + \ddx \sigma^a(\rho_b) 
                                                                                 \\
                                                                                 \label{eq:finalm}
\dot{\rho_b} &=  &-  \ddx (\rho_b \dot{ u}) + D \ddxx \rho_b +  {\cal S}_m(\epsilon, \rho_b) 
\label{eq:finala}
\eea
where $\Phi^{\prime}\equiv \frac{\delta \Phi}{\delta \epsilon}$, $\sigma^a(\rho_b) = \f{-\zeta_1  \Delta \mu \rho_b}{1+\zeta_2 \rho_b}  \chi(\rho_0)$ and 
myosin turnover ${\cal S}_m(\e,\rho_b)=-k_{u0}e^{\alpha \e} \rho_b + k_b \l 1-c\e \r$. 

These equations are numerically solved with either periodic or Neumann boundary conditions using a finite difference scheme ({\it Methods}). Initial conditions
are small amplitude random fluctuations about the homogeneous unstrained state.
The numerical phase diagram, displayed in Fig.\,\ref{fig:numericalPD}, shows several new features compared to the linear phase diagram, 
with direct relevance to experiments. This and a detailed confrontation with experiments are  
 highlighted below.

   \subsubsection{Steady state phase diagram}
   The two  features that are expected to arise from nonlinear effects, namely, the tempering of the linear instabilities to obtain {\it both} finite-amplitude
   oscillatory and  finite-amplitude contractile collapse phases at intermediate and high contractile stresses, respectively, show up in the 
   steady state phase diagram, Fig.\,\ref{fig:numericalPD}a,b. The corresponding kymographs in the bound myosin density (Fig.\,\ref{fig:numericalPD}d,h)
   show the appearance of these steady state at late times. The time development of configurations in these phases can be summarized as follows - starting from 
   a generic state with small random fluctuations about the homogeneous unstrained state, the configuration
 quickly results in a spatially heterogenous (un)binding
 of myosin filaments onto the actin mesh, transiently generating localized compression.
This will increase the local concentration of actin, which in turn will facilitate more myosin recruitment and hence more compression. This local 
compression will cause an elastic tug at the boundaries, enhancing the local strain, which can induce an enhanced myosin unbinding. 
If it does, this will lead to a relaxation of the compressed region, to be followed by another round of binding-compression-unbinding
leading to the observed oscillations. 
  In this spontaneous oscillating phase, the frequency gets smaller with increasing the active stress or decreasing unbinding rate \cite{akankshi}. 
On the other hand, if  myosin unbinding does not occur fast enough, the elastomer will undergo a 
contractile instability, to be eventually stabilised by nonlinear effects such as steric hinderance and filament rigidity. 

In addition to this however, there is a wholly unexpected feature that emerges from a solution of the full nonlinear equations.
In the parameter regime between the oscillatory and the contractile collapse phases, there appears a {\it moving phase} (Fig.\,\ref{fig:numericalPD}), where spatially 
localized actomyosin-dense regions (which we later identify as traveling fronts) spontaneously move to either the left or right boundary. 
In the regimes between the pure moving phase and the oscillatory and collapse phases, lie
the {\it coexistence phases} where the moving phase coexists with oscillations and collapse, respectively.
The corresponding kymographs in the bound myosin density (Fig.\,\ref{fig:numericalPD}e-g)
show the appearance of these steady state at late times. In the {\it SI} we provide a simple explanation based on a time scale argument for the occurrence of these phase transitions.

In making direct quantitative comparisons of the phase transitions predicted by theory with that obtained from experiments at this stage, one should exercise some
 caution, for this requires an understanding of how theoretical quantities such as myosin unbinding rates, mesh stiffness and active contractile stress, depend on experimentally accessible parameters such as phospho-cycling of myosin regulatory light chain \cite{Vasquez_2014,akankshi}, or the strength and density of actin crosslinkers. Typically, altering experimental accessible parameters will affect more than one theoretical parameter, and will thus trace out a 
 trajectory in this theoretical phase diagram. Nevertheless one can make several qualitative assertions based on the theory. Some are immediate, such as
 (i) the existence of bounded (finite-amplitude) oscillations requires {\it both} strain-dependent unbinding and turnover of myosin, (ii)  the coexisting oscillation-moving and collapse-moving phases cannot be obtained in the absence of strain-dependent unbinding, (iii)  advection is a necessary condition for front movement.

We will now explore the relationship between the spontaneous moving phase predicted in this affine theory, 
and the actomyosin flows seen in-vivo, in terms of the shape of the actomyosin-dense region, the onset of movement, the driving force for movement
and its dynamics.
 
\subsubsection{Approach to the flowing steady state - nucleation, growth, coalescence}
In the moving regime, the effective `elastic free-energy' functional $\Phi(\epsilon)$ develops a second minima at $\epsilon= \epsilon_0$ corresponding to
a local compression due to contractility (Fig.\,S3a). 
The initiation of flow starts with the binding of myosin onto regions where the local actin density is high, leading to an eventual 
nucleation of actomyosin-dense regions,
which subsequently grow and coalesce to form larger actomyosin-dense regions. 
This
is best seen
 using a space-time analysis of Eqs.\,\ref{eq:finalu}-\ref{eq:finala} with initial conditions stated above. 

\begin{figure*}
   \centering
     \includegraphics[width=100mm]{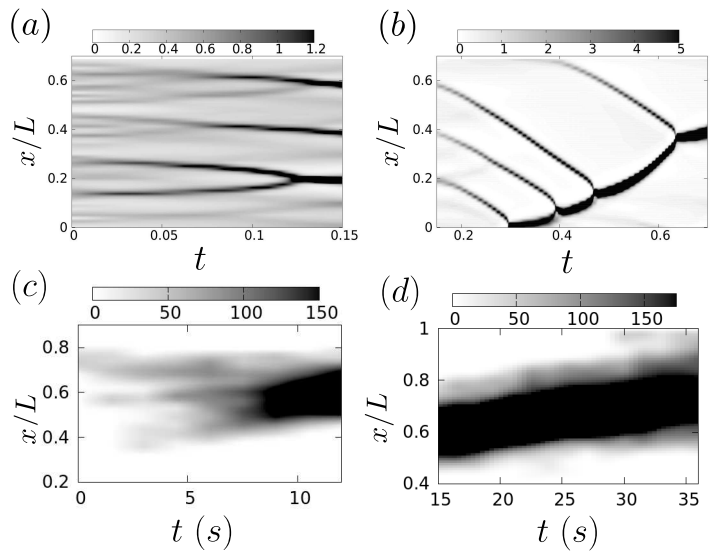}
   \caption{Kymographs in the {\it flow phase} from theory and experiment : (a-b) Kymograph of the spatial profile of bound myosin density from theory, shows nucleation and growth ($0< t< 0.15$),
   followed by coalescence ($0.15< t< 0.7$) and flow ($ t>0.7$). Here $B=6.0$, $-\zeta_1 \m=5.5$, $k=0.5$, $\alpha=1.0$ and $D=0.15$.
            Rest of the parameters as in Fig.\,\ref{fig:numericalPD}.
             (b) Kymograph of the spatial profile of labeled myosin from experiment, shows nucleation and growth ($0< t< 5$s),
   followed by coalescence ($5< t< 10$s) and eventual flow ($ t>10$s).
            \label{fig:kymograph}
	    } 
   \end{figure*}

Kymographs of the spatial profile of bound myosin density, calculated from the theory, show nucleation and growth ($0< t< 0.15$)
   followed by coalescence ($0.15< t< 0.7$) and eventual flow ($ t>0.7$) (Fig.\,\ref{fig:kymograph}a,b). 
   This space-time behaviour accurately recapitulates the dynamics of medial myosin in-vivo as seen from the experimental kymographs, Fig.\,\ref{fig:kymograph}c,d.
   
\subsubsection{Flow starts with the emergence of a compact asymmetric actomyosin profile}
\begin {figure*}
\centering
\includegraphics[width=120mm]{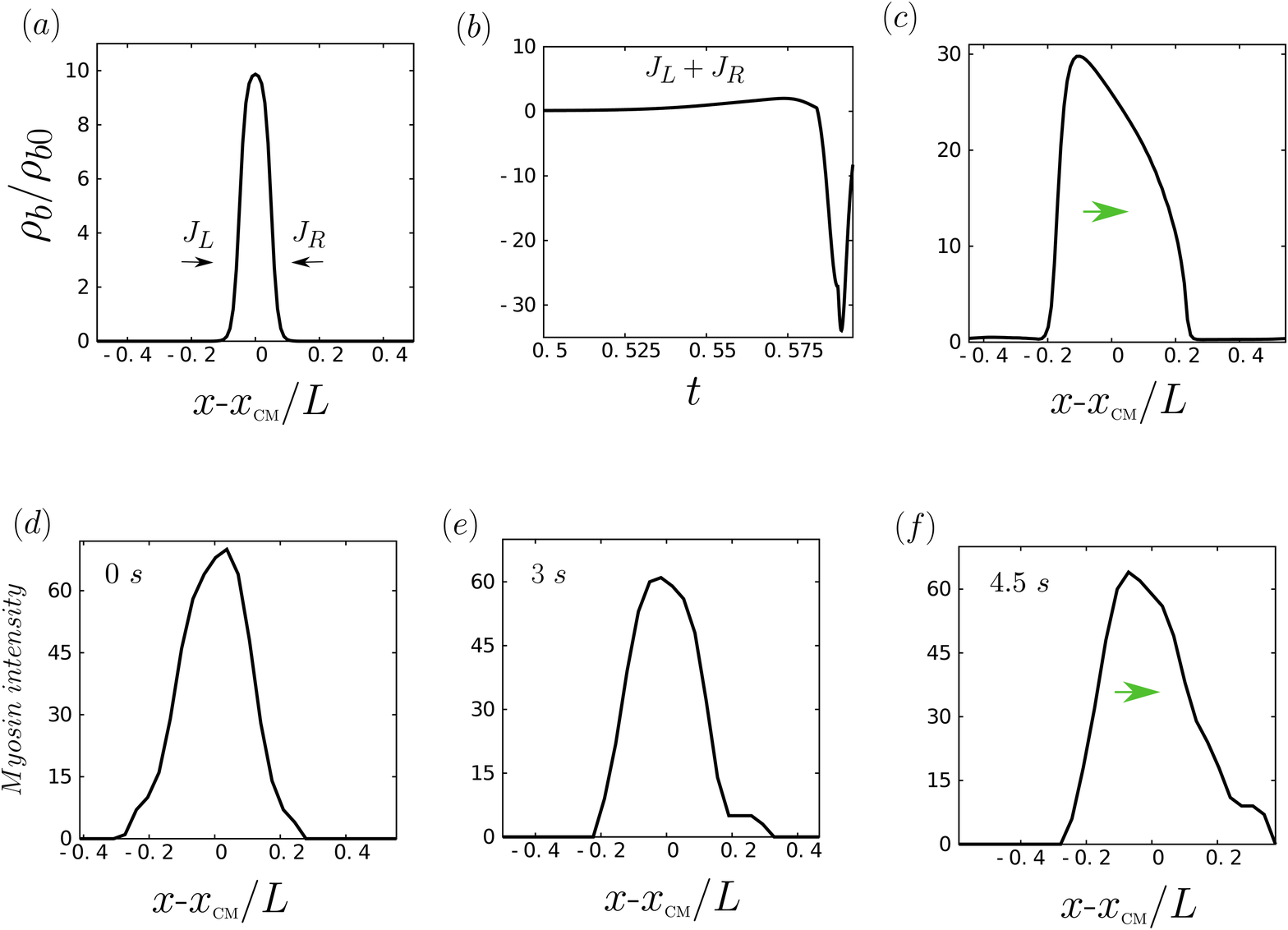}
\caption{Time evolution of a stationary symmetric profile of $\rho_b$ to an asymmetric traveling front : (a) Symmetric myosin density profile, about which we compute the instantaneous 
left-right fluxes, $J_{L,R}$, symmetrically situated about the initial profile. (b) Time evolution of the algebraic sum $J_L + J_R$, shows a net flux in one direction
(here, from the right) as a precursor to the asymmetric traveling front.  (c) Emergence of the asymmetric traveling front which moves towards the right in a shape-invariant manner.
(d,f) An example from experiments which shows how an initial stationary symmetric myosin profile at $t=0$, evolves to $t=3.0$s, and finally to an asymmetric profile at $t=4.5$s that travels to the right.
	  }
	  \label{fig:flux}
\end{figure*}

What prompts the actomyosin-dense region to flow ? We study the configuration of the localized actomyosin-dense region
before the flow commences, and find that it assumes  a symmetric localized profile (Fig.\,\ref{fig:flux}a) within which the 
strain $\epsilon=\epsilon_0$ (the second minimum) where $\Phi^{\prime}(\epsilon)=0$. The active stress in this region
is higher than out, this gradient in stress should
induce inflowing myosin currents from either side of it. We verify this by monitoring
the fluxes $J_{L}$ and $J_R$,  coming from the left and right of
this symmetric profile. Over time, owing to stochasticity either  in the initial conditions or the dynamics, 
there is a net flux, $J_L+J_R$, from either the left or the right (Fig.\,\ref{fig:flux}b), leading an asymmetric profile (Fig.\,\ref{fig:flux}c),
and hence gradient of $\rho_b$ across the profile. This marks the onset of the traveling front. This exactly replicates the situation
in-vivo, as seen in Fig.\,\ref{fig:flux}d-f.

\begin{figure*}
   \centering
       \includegraphics[width=150mm]{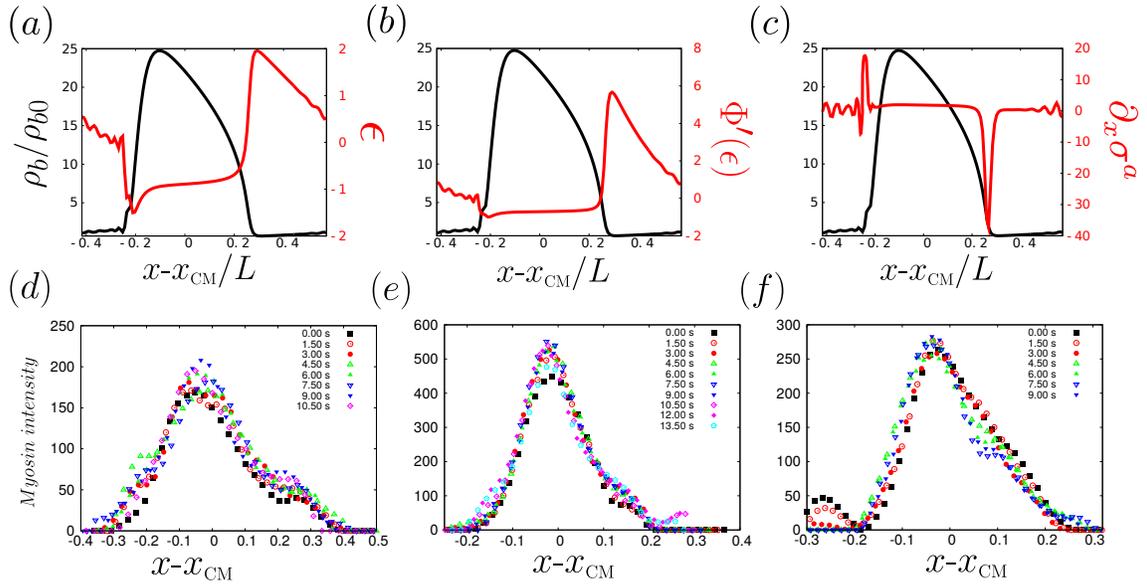}

    \caption{Anatomy of the traveling front in the co-moving frame : (a) Spatial profile of Excess bound myosin density (black) and strain $\epsilon$ (red) profile, (b) Spatial profile of Myosin density and derivative of the effective elastic free energy
    $\Phi^{\prime}(\epsilon)$ (red), 
            (c) Spatial profile of myosin density and active force (red). Horizontal axis is distance from centre of mass position, $x_{CM}$. Here $B=6.0$, $-\zeta_1 \m=4.8$, $k=0.2$, 
            $\alpha=1.0$ and $D=0.25$.     (d-f) Three separate examples of the spatial profile of myosin intensity in the comoving frame of the traveling front displayed at different times. Note that the traveling front hardly changes its shape as it flows towards a junction.
	 \label{fig:anatomy}
 	    } 
   \end{figure*}

\subsubsection{Anatomy of the traveling front}

Our theory predicts that  the myosin-dense cluster starts to flow only when it has attained an asymmetric density profile, whereupon
it moves as a traveling front, with a constant velocity while maintaining its shape (as long as there are no further coalescence events). We confirm this using a 
variety of initial conditions or the bound myosin density $\rho_b$, including starting with a single symmetric gaussian profile.
We analyze the asymmetric profile of the traveling front
by transforming to the co-moving frame $x\pm vt$ (Fig.\,\ref{fig:anatomy}).

We find that within the traveling front, the strain takes a value slightly more compressed relative to $\epsilon_0$, the value of the strain at the second minima,
where $\Phi^{\prime}=0$. The traveling front is stably compressed in a force-free state (Fig.\,\ref{fig:anatomy}a,b). The asymmetric myosin profile 
gives rise to a gradient in the active stress (Fig.\,\ref{fig:anatomy}c), which provides the propulsive force for the traveling front to move to the right in
Fig.\,\ref{fig:anatomy}c.

The fact that the flowing actomyosin-dense region commences to move when it has attained an asymmetric profile, moves with a constant velocity in the direction
where the leading edge has the smoother slope and
maintains its asymmetric shape as it moves, exactly replicates the situation in-vivo, as seen in Fig.\,\ref{fig:anatomy}d-f.

\subsubsection{Driving force for flow is established within the medial actomyosin-dense cluster and not the cell boundary}
\begin{figure*}
   \centering
       \includegraphics[width=120mm]{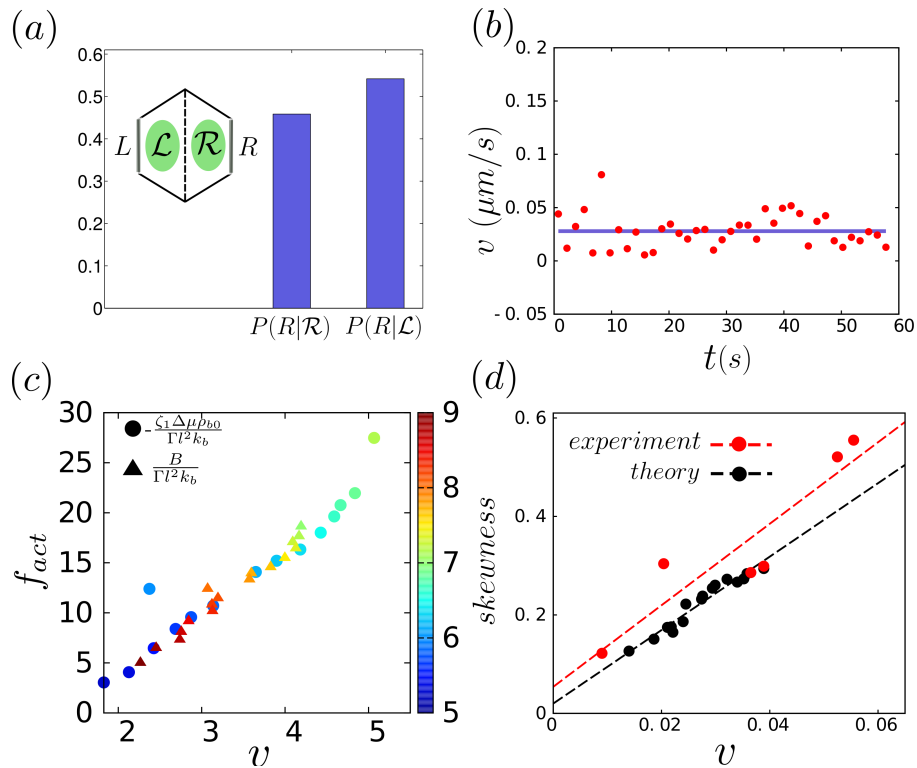}
   \caption{Velocity and driving force of traveling front : (a) Histogram of the number of flows that move to the right junction $R$ starting from either the right ${\cal R}$
($P(R \vert {\cal R}$)  or the left  ${\cal L}$ ($P(R \vert {\cal L}$)  region of the cell (inset shows schematic). Data collected from $24$ actomyosin-dense clusters over $18$ cells. The fact that the histograms are similar is consistent with the theoretical prediction that the flow is spontaneous and not driven by the cell boundaries. (b) Velocity of an isolated flowing actomyosin-dense cluster monitored over time shows that it is a constant, as predicted by theory. (c) Theory predicts that the traveling front velocity is proportional to the net active force integrated over the front profile across the moving front. We demonstrate this fact from a numerical solution of the dynamical equations by varying the parameters of the active stress (circle) and the elastic stress (triangle). for different values of $B$ (with $-\zeta_1 \m=6.0$, $k=0.2$ and $D=0.15$ fixed) and $-\zeta_1 \m$ (with $B=8.0$, $k=0.2$ and $D=0.15$ fixed).
                    Rest of the parameters as in Fig.\,\ref{fig:numericalPD}.
The colour bar shows the magnitude of these stresses in dimensionless units. (d) Theory predicts that the {\it skewness} of the traveling front profile is proportional
to the velocity (black circles). We verify this relationship from an analysis of the myosin intensity images (red circle) across $6$ actomyosin-dense clusters and $4  $ cells.
We have chosen a pair of points from theory and experiment which have the same value of skewness and have scaled their respective velocities to be the same. We have used this same scaling for all the other points to obtain this graph.
	 \label{fig:drivingforce}
	    } 
   \end{figure*}

The dynamics of the traveling front that emerges from our theory is local, its propulsion is therefore independent of the boundary or the distance from the 
boundary. We ask whether this is true of the flowing actomyosin-dense regions in-vivo. Figure\,\ref{fig:drivingforce}a shows that irrespective of its initial position  at the commencement of the flow, the moving actomyosin-dense region travels to the left or right cell boundary with equal probability. Further, Fig.\,\ref{fig:drivingforce}b
shows that the moving actomyosin-dense region travels with a constant velocity as it moves towards a given cell boundary, its speed does not depend on the distance
from the cell boundary.

To emphasize this point, we calculate the velocity of the traveling front  by
 integrating Eqs.\,(\ref{eq:finalu}-\ref{eq:finalm}) across the scale ${\Omega}$ of the traveling front in the co-moving frame. This leads to the formula,
$v = \Gamma^{-1} \int_{\Omega} \partial_x \sigma^a \equiv \Gamma^{-1} f_{act}$, which states that the velocity depends only on the shape asymmetry of the front; if the 
shape is maintained over time, then the velocity is a constant. In Fig.\,\ref{fig:drivingforce}c, we plot the $f_{act}$ versus the velocity and show that they are proportional 
to each other over a large range of active and elastic stresses. Does this relation hold in the in-vivo context ? Rather than extracting $f_{act}$ from the myosin intensity,
we compute the shape-asymmetry via the skewness, $S\equiv \f{\int_{\Omega} \l x - x_{CM}\r^3 \rho_b(x)}{\l \int_{\Omega} \l x - x_{CM}\r^2 \rho_b(x)\r^{3/2}}$, of the
myosin profile. The comparison between the theoretical prediction and in-vivo experiment is striking (Fig.\,\ref{fig:drivingforce}d).

To our mind, this establishes unambiguously that the flow towards the junctions is spontaneous with the driving force coming from the gradient in myosin 
established within the front. The boundary does not affect the flow speed, at best,  weak asymmetries that may arise at the boundary (for instance due to
an asymmetry in functional cadherin) may bias the direction of the flow \cite{Lecuit2013}. 

One consequence of this is that a traveling front moving to the right, might reverse its direction following a coalescence with another traveling front 
(Fig.\,S5a). Figure\,S5b-d  shows examples of such reversals observed in in-vivo experiments.

\subsection{Nonaffine deformations and intranetwork flow}
\begin{figure*}
   \centering
         \includegraphics[width=120mm]{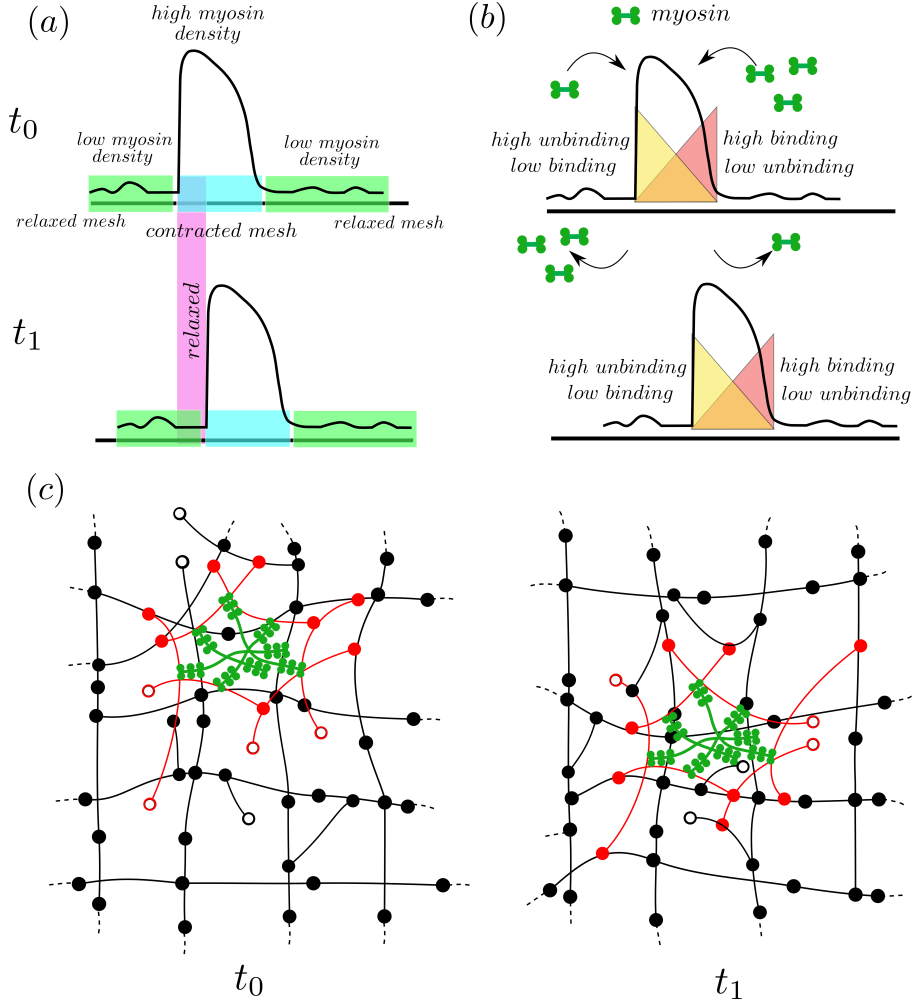}
   \caption{Possible mechanisms for the movement of actomyosin-dense structures in an active elastomer : Affine deformations - (a) moving deformation of the actomyosin mesh without turnover, implying a traveling front, and (b)  moving deformation of the actomyosin mesh with differential myosin binding unbind rates at the leading and trailing edges of the front.
   Nonaffine deformations - (c) Schematic showing the movement of the actomyosin-dense region in a  transient actomyosin network. Filled (open) circles are bound (unbound) crosslinkers, myosin minifilaments in green, and black lines are actin filaments comprising the background mesh
   and red lines are actin filaments comprising the actomyosin dense region. The schematic depicts myosin induced intranetwork flow.
	 \label{fig:massflow}
	    } 
   \end{figure*}

It is important to note that the movement of the actomyosin-dense region arising from affine deformations of the active elastomer
is a moving {\it deformation} of the actomyosin mesh, and once established, is not contingent on myosin turnover, as shown in Fig\,\ref{fig:massflow}a. 
One could also sustain a traveling front or moving deformation of the actomyosin mesh by ensuring differential myosin binding unbind rates at the leading and trailing 
edges of the front, in a kind of {\it treadmilling} movement, Fig\,\ref{fig:massflow}b. None of these however is associated with mass flow of actin and myosin.

Preliminary experiments  show loss of recovery on performing a FRAP on a small region 
within the actomyosin-dense cluster \cite{Lecuit2010}, suggesting the possibility of actual mass flow. To realise this within the active elastomer description,
we should first recognise that when the local 
contractile stresses are large, the deformation on the mesh may no longer be affine, and one needs a description of {\it nonaffine} deformations of the active elastomer. 

The following picture helps understand the origins of nonaffineness.  Consider a disordered mesh comprising actin filaments linked to each other by crosslinkers   such as $\alpha$-actinin and myosin (we will assume that this is an {\it unentangled} network). The bound myosin locally compresses the mesh here and there, recruiting more myosin in the 
process. When the local bound myosin concentration goes beyond a threshold (so that the configuration now samples the second minimum of the effective free energy, $\Phi(\epsilon)$),
the local compression is high and the regions surrounding this myosin-dense region get significantly stretched. This could lead to a breaking (or ripping) of the mesh,
either via the unbinding of crosslinkers or by the sliding and slipping of filaments past each other. Within the active elastomer framework, this can be accounted for by 
turnover of actin and crosslinkers. This mesh breakage subsequently heals by the rebinding of the crosslinkers or actin itself.

With this in mind, we refer to several seminal studies on
thermally activated reversibly crosslinked networks in the context of the dynamic properties of {\it physical gels} \cite{muthukumar,leibler,tanakaedwards}.
The most dramatic feature of such reversible networks is its {\it internal fluidity}, where each chain can diffuse through the entire network due to the finiteness of the crosslinker life time, in spite of being partially connected to the macroscopic network structure in the course of movement. These systems thus flow under an external stress on time scales longer 
 than the crosslinker dissociation time (see, \cite{munro} in the context of actin mesh). 

In the present context, when the local bound myosin concentration rises beyond a threshold and attains an asymmetric density profile, it induces mesh breakage in its surrounding regions. This
actomyosin-dense region can move through the entire network due to the finiteness of the crosslinker life time, in spite of being partially connected to the macroscopic network structure in the course of movement. These systems should thus exhibit {\it flow} under an internal active contractile stresses on time scales longer 
 than the crosslinker dissociation time. 
This is depicted in Fig\,\ref{fig:massflow}c. 

To describe this mathematically, it is convenient to define physical quantities coarse-grained over the scale of the actomyosin-dense region $\Omega$, such as
$\bar{\rho}_b = \Omega^{-1} \int_{\Omega} \rho_b$ and ${\bf p} = \Omega^{-1} \int_{\Omega}  \boldsymbol \nabla \cdot \boldsymbol \sigma^a$, the net force-dipole associated with
the anisotropy of the myosin profile (note we now have to work in $d\geq2$). We may decompose myosin density configuration
 $\bar{\rho}_b$ into a sum of actomyosin-dense clumps (contributing to nonaffine deformations) and a background (contributing to affine deformations), with volume fractions
$\phi$ and $1-\phi$, respectively. The equation for the $\bar{\rho}_b$ may now be written as,
\be
\dot{\bar{\rho_b}} =  -   \boldsymbol \nabla \cdot \l \bar{\rho}_b {\bf v} \r + D (1-\phi) \nabla^2 \bar{\rho}_b
+  {\cal S}_m(\bar{\rho}_b) 
\label{eq:omegaav}
\ee
where, ${\bf v} = \phi \beta \l {\bf p} + A {\bf p} \cdot \boldsymbol \epsilon\r + (1-\phi) {\dot {\bf u}}$, and $\beta$ is the mesh breakage probability and $A$ is a proportionality
constant.

Flow described here is a consequence of internal active deformations in a
transient actomyosin network. We believe this is  a new physical phenomenon, an active version of the flow observed in physical gels under external load. 
A more complete theory of nonaffine deformations of a transient actomyosin network is a task for the future.

\section{Discussion}
 Although highly simplified, in that it has
completely ignored the coupling of actomyosin dynamics to local chemical signalling such as Rho \cite{akankshi},
we believe this active elastomer model with strain-dependent turnover of components, admitting {\it both} affine and nonaffine deformations,
captures the essential physics of actomyosin pulsation and flows 
observed in a wide variety of tissue remodelling contexts such as {\it Drosophila} germ band extension and dorsal closure in the amnioserosa.
The minimal ingredients for actomyosin pulsation and flow are mesh-elasticity, actomyosin contractility, advection and turnover of both myosin and actin.
While sustained oscillations would be expected from a model of 
an active elastomer, the occurrence of spontaneous, symmetry breaking flows via a shape-preserving profile is rather unexpected.

In this study, we have modelled the medial actin mesh during apical constriction and germ band extension in the {\it Drosophila} embryo 
as an active elastomer embedded in a solvent \cite{Lecuit2013,Martin_jcb2010,rohjohnson,Wu_NCB_2014}, rather than as an active fluid, as has been successfully done
in the context of {\it C. Elegans} embryo \cite{Mayer_Nature_2010,grill1}.
There are several empirical reasons for doing this - (i) the pulsation of medial actomyosin in these systems is correlated with the oscillations in the area of the apical surface
\cite{Martin,akankshi,zallen}, (ii) Modulating the strength of the coupling of medial actin to the junction via $\beta$-catenin affects the pulsation, and (iii) turnover time of actin is not small.

To make a detailed comparison of the spatiotemporal actomyosin patterns with experiments generated using quantitative imaging, we will need 
to extend this numerical study to 2-dimensions, using appropriate (anisotropic) boundary conditions and allowing for shear.
The nucleation and growth of the actomyosin-rich domains are similar to that seen in 1-dim, with the difference being that domains can move around each other in 2dim 
and can exhibit anisotropic movement.

In future, we would like to extend this framework to understand the dynamical coupling of the medial actomyosin with degrees of freedom (concentration of E-cadherin) attached to a deformable cell junction. Though the emergence of actomyosin flows does not depend on specific boundary conditions, cell boundaries may directionally bias the intrinsic ability of actomyosin networks to generate flow, as proposed before \cite{Lecuit2013}.

The authors would like to thank the CNRS (Laboratoire International Associ\'e between IBDM and NCBS), AMIDEX and the 
Labex INFORM (ANR-11-LABX-0054), for support. AM and TL were supported by the ERC (Biomecamorph 323027).
DB and MR thank K. Vijaykumar, S. Ramaswamy and R. Morris for extremely useful discussions.

\noindent
{\bf Methods}\\
\small{\bf Free-energy functional}\\
The bulk elastic free-energy density of the elastomer in terms of the linearized strain tensor $\epsilon_{ij}$ is given by
$f_B = \frac{\mu}{2} \epsilon_{ij}\epsilon_{ij} + \frac{\lambda}{2} \epsilon_{ii}\epsilon_{jj}+ \ldots + C \delta \rho \epsilon_{ii} + \frac{A}{2} \delta \rho^2 + \ldots$, where the $\ldots$ represent higher order terms to 
stabilize a possible activity induced contracted state and to prevent a runway density increase ($\delta \rho$ is the deviation of the local mesh density from its equilibrium value,
$\delta \rho = -(C/A)\epsilon_{ii}$).

\noindent
\small{\bf Parameter values}\\
Here we relate the values of the dimensionless parameters to real units extracted from a variety of experimental measurements.
For the unit of length, $l=\sqrt{\eta/\Gamma}$ or the actin mesh size, we take $0.5\mu$m (consistent with the rough estimates in \cite{Lecuit2010})
The unit of time, $k_b^{-1}$, can be estimated from the myosin FRAP data \cite{akankshi} - we find $k_u=0.2$\,s$^{-1}$, and taking the ratio
$k=k_{u}/k_b$ to be $0.1$ (consistent with simulation studies \cite{kamm} and experiments \cite{murrell}), we obtain a binding rate,
 $k_b=2.0$\,s$^{-1}$.
The cytosolic viscosity in germ band cells has been measured  by micro-rheology to be $3.6\pm0.1$\,Pa.s \cite{lenne_PNAS15}.

We can now convert all the dimensionless values into real values, and check for consistency with other experimental estimates.
Thus, a dimensionless value of the bulk modulus $B=5$ translates to $B=36.5$\,Pa (consistent with what can be estimated from \cite{lenne_PNAS15}).
Similarly, a dimensionless value of the magnitude of the active stress, $\vert \zeta_1\m\vert=5$ translates to $\vert \zeta_1\m\vert =36$\,Pa (roughly the order
of magnitude estimated from  \cite{lenne_PNAS15}). Finally, the dimensionless diffusion coefficient $D=0.25$ implies a real value of $D=0.12\mu$m$^2$.

This implies that a dimensionless front velocity $v=1$ (see, Fig.\,\ref{fig:drivingforce}c) translates to a real velocity of $v=1.0\mu$m/s.
\\

\noindent
\small{\bf Numerical methods}\\
Since the dynamical equations have nonlinear advection and diffusion, care must be taken
in evaluating the flux due to dissipative and dispersive errors arising from spatial discretization.
We use the finite volume method for spatial discretization \cite{Sweby1984},
which has been found to be useful for non-linear advection equations \cite{leveque}. 

We calculate the numerical flux using the Van-Leer's flux limiter, 
which uses a different formula to calculate the spatial derivative depending on how sharply the $\rho_b$ profile
changes in space. When the profile changes very fast, the scheme implements the upwind method \cite{Courant1952} which reduces the dispersion error
through numerical diffusion. When the profile changes smoothly the scheme implements a second order accurate method called Lax-Wendroff method
\cite{Lax1960}.
In our numerical scheme, the density ($\rho_b$ or $\rho$) flux  on the interface between $i^{th}$ and $(i-1)^{th}$ node is computed as,
\be
\begin{split}
f^{n+\f{1}{2}}_{i-\f{1}{2}} = \f{1}{2} v_{i-\f{1}{2}} [(1+\theta_{i-\f{1}{2}})\rho^{n}_{i-1} + (1-\theta_{i-\f{1}{2}})\rho^{n}_{i}] \\
			      + \f{1}{2} \vert v_{i-\f{1}{2}} \vert \l 1-\Bigg\vert\f{v_{i-\f{1}{2}}\Delta t}{\Delta x}\Bigg\vert\r 
			        \phi^n_{i-\f{1}{2}}(r^n_{i-\f{1}{2}}) \l \rho^n_i - \rho^n_{i-1} \r
\end{split}
\ee
Here
$v_{i-\f{1}{2}} = \f{v_i - v_{i-1}}{2}$ and 
\be
\theta_{i-\f{1}{2}} = 
\begin{cases}
+1, & \text{if }v_{i-\f{1}{2}}>0 \\
-1, & \text{if }v_{i-\f{1}{2}}\leq0
\end{cases}
\ee


\be
r^n_{i-\f{1}{2}} = 
\begin{cases}
\f{\rho^n_{i-1} - \rho^n_{i-2}}{\rho^n_i - \rho^n_{i-1}}, & \text{if }v_{i-\f{1}{2}}>0 \\
\f{\rho^n_{i+1} - \rho^n_i}{\rho^n_i - \rho^n_{i-1}}, & \text{if }v_{i-\f{1}{2}}\leq0
\end{cases}
\ee
The function $\phi(r)$ is the Van Leer flux limiter  
\be
\phi(r) = \f{r+ \vert r \vert}{1+\vert r \vert} \, .
\ee
The time integration is done with a total variation diminishing (TVD) $3^{rd}$order Runge-kutta method \cite{tvdRK}. All other derivative terms 
were
discretized using a simple finite difference.
The initial conditions were chosen from a uniform random distribution of fixed width about a uniform, unstrained configuration.
We used periodic boundary conditions throughout and a time-space discretization, $\Delta t=10^{-4}$ and $\Delta x=5\times10^{-2}$. 

\noindent
\small{\bf Experimental methods}\\
Embryos were prepared as described before \cite{Cavey&Lecuit}. Nikon spinning disc Eclipse Ti inverted microscope using 100X, 1.4 N.A oil immersion objective was used to do time lapse imaging from during stage 7. The system acquires images using the MetaMorph software. Starting from the most apical plane, 4-7 z-sections $0.5\,\mu$m
 apart were acquired every $1.5-4$s  (depending on the experiment) using a single camera. Sum-intensity z-projection of slices was used for all quantifications, followed by a background subtraction using the available plugin in Fiji.

\clearpage


\begin{center}
\large{Supplementary Online Material for}
\vskip 0.2in
\Large{\bf Actomyosin pulsation and symmetry breaking flows in a confined active elastomer}
\vskip 0.2in
\Large{Deb Sankar Banerjee$^{1}$, Akankshi Munjal$^{2}$, Thomas Lecuit$^{2}$ and Madan Rao$^{1,3,\ast}$\\
\vskip 0.1in
\normalsize{$^{1}$Raman Research Institute, C.V. Raman Avenue, Bangalore 560080, India}\\
\normalsize{$^{2}$IBDML, CNRS-Universite de la Mediterranee, France}\\
\normalsize{$^{3}$Simons Centre for the Study of Living Machines, National Centre for Biological Sciences (TIFR), Bellary Road, Bangalore 560065, India}\\
\vskip 0.1in
\normalsize{$^\ast$To whom correspondence should be addressed; E-mail:  madan@ncbs.res.in.}
}

\end{center}


\section*{Hydrodynamic equations of an Active Elastomer}

The hydrodynamics of an active elastomer embedded in a fluid solvent is described by the hydrodynamic variables - 
(i) density of filamentous
mesh, $\rho$; (ii) displacement field of filamentous mesh, $\bs u$; (iii) density of 
bound myosin minifilaments, ${\rho_b}$, and (iv) fluid (solvent) velocity ${\bs v}$. 
The  linearized elastic strain is defined as $\e_{ij} = 1/2 (\partial_i u_j + \partial_j u_i)$. Our treatment closely follows \cite{Marchetti2011a,Marchetti2011b}.

The hydrodynamic equations of the active elastomer mesh in bulk are given by,
\begin{equation}
\rho\ddot{\bs u}+\Gamma ({\dot {\bs u}-{\bs v}})  = \bs \nabla \cdot (\bs \sigma^e + \bs \sigma^a 
                                                                           +\bs \sigma^{d} )
\label{eq:mesh}
\end{equation}
\begin{equation}
\dot{\rho_b} + \bs \nabla \cdot (\rho_b \dot{\bs u})= D\bs\nabla^2\rho_b + {\cal S}_m
\label{eq:myosin}
\end{equation}
\begin{equation}
\dot{\rho} + \boldsymbol \nabla \cdot (\rho \dot{\boldsymbol u}) =  {\cal M} \nabla^2 \frac{\delta F}{\delta \rho} + {\cal S}_a
\label{eq:actin}
\end{equation}
where  ${\bs v}$ is the fluid velocity and the rest of the parameters have been defined in the main text.
The friction $\Gamma$ between the mesh and cytosol can in principle depend on the mesh density. 
The dynamics of  both actin filaments and myosin minifilaments show a  turnover over a scale of minutes, which we 
refer to as ${\cal S}_a$ and ${\cal S}_m$, respectively.

The constitutive relations for the elastic ($\sigma^e$) and dissipative ($\sigma^d$) stresses are,
\bea
\sigma^e_{ij} & = & \l\lambda+\f{2\mu}{3}\r \delta_{ij} {\bs \d} \cdot {\bs u} + 2\mu \l \e_{ij} - \f{1}{3}\delta_{ij} {\bs \d} \cdot {\bs u} \r     \\
\sigma^d_{ij} & = & \eta_b \delta_{ij} {\bs \d} \cdot {\dot {\bs u}}  + 2\eta_s \l \dot{\e}_{ij}-\f{1}{3}\delta_{ij}{\bs \d} \cdot {\dot {\bs u}} \r
\label{eq:constitutive}
\eea
Here $\lambda$, $\mu$ are the Lam\'e coefficients of the elastic mesh and $\eta_b$, $\eta_s$ are the bulk and shear viscosities of the mesh, respectively. 
The form of the active stress $\sigma^a$ is described in the main text.

The fluid hydrodynamics is given by,
\begin{equation}
\rho_f (\dot{\bs v}+{\bs v} \cdot {\bs \d} \bs v) = \eta^s_f {\bs\d}^2 {\bs v} +(\f{\eta^s_f}{3}+\eta_f^b){\bs\d}({\bs \d \cdot \bs v}) - {\bs\d} P + \Gamma ({\dot {\bs u}-{\bs v}}) 
\label{eq:fluid} 
\end{equation}
where $\rho_f$ is the density of the fluid and $\eta^s_f$ and $\eta_f^b$ are the shear and bulk viscosities of the fluid.

Finally, the pressure $P$ is eliminated by demanding total incompressibility,
\be
{\bs \d} \cdot \l (1-\phi){\bs v} + \phi {\dot {\bs u}} \r = 0
\label{eq:incompress}
\ee
where $\phi$ denotes the volume fraction of the actomyosin mesh. For a gel, we expect $\phi \ll 1$,
yielding, ${\bs \d} \cdot {\bs v}  = 0$.

These are the complete set of equations, which we display here for completeness.

Since the cytosol is a low Reynolds number fluid, Eq.\,(\ref{eq:fluid}) reduces to a force balance condition, where the right hand side of the
equation equals zero. In the main manuscript, we study the Rouse limit where we ignore the hydrodynamics of the fluid ${\bs v}$;
this is suggested by our experiments that show the actin mesh moves with respect to the fluid, and does not carry (advect) the fluid and other soluble 
molecules along with it, except for those which are bound to the mesh \cite{akankshi,Lecuit2010}. 

Since we are interested in the
over-damped dynamics of the mesh, we ignore inertia and drop the $\ddot{\bs u}$ term in Eq.\,(\ref{eq:mesh}). This leads to
Eqs.\,(1), (2) and (3) in the main text.

\section*{Linear Stability Analysis}

For the linear analysis, we only consider Eqs.\,(\ref{eq:mesh}) and (\ref{eq:myosin}) with the mesh density slaved to the elastic
compression $\epsilon_{ii}$ (as described in the main text).
Equations (\ref{eq:mesh}) and (\ref{eq:myosin}), after going to the Rouse and overdamped limit, look like
\be
\Gamma \dot{\bs{u}} = \bs\nabla\cdot({\bs{\sigma}}^e + {\bs{\sigma}}^a + {\bs{\sigma}}^d)
\ee
and 
\be 
\dot{\rho_b} + \bs\nabla\cdot{(\rho_b \dot{\bs u})} =  D\bs\nabla^2\rho_b - k_{u0} e^{\alpha \bs\nabla\cdot{\bs u}} \rho_b + k_b \rho
\ee
where we have unpacked the myosin turnover $ {\cal S}_m= -k_u(\epsilon) \rho_b + k_b \rho$. For this linear analysis, we take the
${\bs{\sigma}}^a= -\zeta_1 \l 1 + \zeta^{\prime} \rho \r  \m \rho_b$, 
${\bs{\sigma}}^e= B \bs{\nabla}\cdot \bs{u}$ and ${\bs{\sigma}}^d= \eta \bs{\nabla}\cdot\dot{\bs{u}} $ , where $\zeta_1 <0$ denotes contractile 
stress, and in this one-constants approximation, $B$ and $\eta$ are given by $\lambda+2\mu$ and $\eta_b+\f{4}{3}\eta_s$, respectively. 
    For convenience, we set $\zeta^{\prime} =1$.

These equations can be rewritten in dimensionless form with time ($t$) and space ($\bs x$) in units of $k_b^{-1}$ and  
$l = \sqrt{\f{\eta}{\Gamma}}$, respectively, leading to the redefinitions,
\bea
 \f{u}{l} \to u	\nn \\
 \rho_b/\rho_{b0} \to \rho_b	\nn \\
 \f{B}{\Gamma k_b l^2} \to B	\nn \\
 \f{\zeta_1\m\rho_{b0}}{\Gamma k_b l^2} \to \zeta_1\m	\nn \\
 \f{k_{u0}}{k_b} \to k \nn \\
 \f{D}{k_b l^2} \to D 
\eea
and the following equations in  dimensionless form
\be
(1-\nabla^2)\dot{\bs{u}} = (B+\zeta_1\m) \nabla^2 \bs u - \zeta_1 \m \bs\nabla \rho_b \nn
\ee
and 
\be 
\dot \rho_b + \bs\nabla\cdot{(\rho_b \dot{\bs u})} = D\bs\nabla^2\rho_b - k e^{\alpha \bs\nabla\cdot{\bs u}} \rho_b + \rho \,\,.\nn
\ee
Upon linearizing about the homogeneous, unstrained fixed point $\l u_0, \rho_{b0}, \rho_0\r$, we obtain,
\be
(1-\nabla^2) \delta \dot{\bs u} = (B+\zeta_1\m) \nabla^2 \delta \bs{u} - {\zeta_1}{\m} {\nabla}{\delta\rho_b} \nn
\ee
and 
\be 
\delta \dot{\rho_b} + \nabla \cdot \delta \dot{\bs u} = D\nabla^2\delta\rho_b - k(\a+c)  \bs\nabla \cdot \delta \bs u  - k \delta \rho_b  \nn
\ee
where we have used the fact that the fluctuation in $\rho$ is slaved to the compression, $\delta \rho = c \epsilon_{ii}$


On Fourier transforming the above equations in space, $f(\bs q,t)= \int_{-\infty}^{\infty} {f(\bs x,t) e^{-i \bs q\cdot \bs x} d\bs x} $, we obtain the matrix equation,
\begin{widetext}
\be
\quad
\begin{bmatrix}
 \delta \dot{u} \\
 \delta \dot{\rho_b}
\end{bmatrix}
=
\begin{bmatrix}
 -\l\f{q^2}{1+q^2}\r (B+\zeta_1\m)  & -\l\f{i q}{1+q^2}\r \zeta_1\m \\
 i q\l \f{q^2}{1+q^2}(B+\zeta_1\m)-(\a + c)k \r  &  -\l k+\f{q^2}{1+q^2} \zeta_1\m + q^2D \r
\end{bmatrix}
\times
\begin{bmatrix}
 \delta u \\
 \delta \rho_b
\end{bmatrix}
\quad
\label{eq:LSmatrix}
\ee
\end{widetext}

Solving (\ref{eq:LSmatrix}) for the two eigenvalues, $\lambda_+$ and $\lambda_-$, we obain the general solution,
\be
u(\bs q,t) = u_1(\bs q) e^{\lambda_+ t} + u_2(\bs q) e^{\lambda_- t}
\ee
where, $\lambda_{\pm} = \lambda_1 \pm \sqrt{\lambda_2}$, with
\begin{widetext}
{\scriptsize
\bea
&&\lambda_1 = \f{-\l k(1+q^2) + Dq^2(1+q^2) + q^2(B+2\zeta_1 \m)\r}{2(1+q^2)} \nn \\
&&\lambda_2 = \f{\l k \l 1+q^2 \r +q^2 \l (B+D+D q^2\r +2 q^2 \zeta_1 \m \r {}^2-4 q^2 \l 1+q^2 \r \l B \l k+D q^2\r + \l D q^2+k (1+c+\alpha )\r \zeta_1 \m \r}{4(1+q^2)^2} 
\label{eq:LSroots}
\eea
}
\end{widetext}


\begin {figure*}
\centering
\includegraphics[width=100mm]{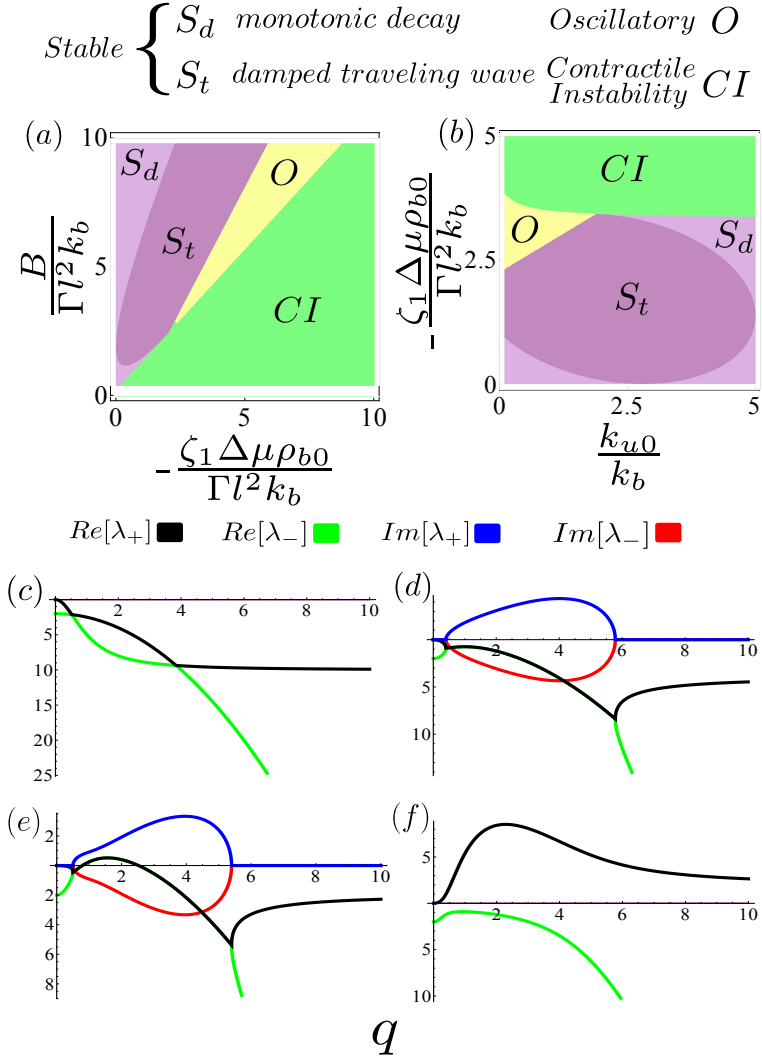}
\caption{(a-b) Linear stability phase diagrams in (a) effective elastic stress density vs.\,contractile stress density at $k=1.0$ and 
  (b)  Effective contractile stress density vs.\,inverse lifetime of bound myosin at $B=4.0$. The stresses are normalized by the frictional stress density, $\Gamma k_b l^2$.
The phases are described in the legend.
Rest of the parameters are $\alpha=0.1, c=0.1, D=0.1$.
(c-f) Typical dispersion curves obtained from linear stability analysis, showing the complex roots $\lambda_{\pm}$ as a function of wave-vector $q$. Colour code displayed above. Panels shows typical behaviour in the (c) stable, (d) damped oscillations, (e) unstable oscillations, and (f) contractile instability phases.
	  }
	  \label{fig:dispersion}
\end{figure*}

From the dispersion relations, we determine 4 phases (a) stable ($Im [\lambda_{\pm}]=0$), (b) damped oscillations, (c) unstable oscillations, and (d) contractile instability ($Im [\lambda_{\pm}]=0$), corresponding to the dispersion curves shown in Fig.\,\ref{fig:dispersion}. The phase boundary in Fig.\,1 of the main text is obtained
by specifying a particular value of $q$ (we have taken $q=2$), or by restricting $q$ to lie between $\[q_{min}, q_{max}\]$, where $q_{min}=2\pi/L$ and   $q_{max}=2\pi/l$ ($=2\pi$, in rescaled units).

The phase boundary between the {\it stable and damped traveling wave phases} can be obtained from the solution of $Im \[\lambda_{\pm}\]=0$,
\be
\zeta_1\m \Big\vert_{S_d \to S_t} = \frac{1}{2} \left(-B - \sqrt{2 B k-k^2+2 B D q^2-2 D k q^2-D^2 q^4}\right)		\nn
\ee

The {\it stable phase} crosses over to the {\it unstable phase}, when
\be
-\l B + 2\zeta_1\m + k + D q_c^2 \r \geq 0 
\ee
(obtained from $Re \[\lambda_{\pm}\]=0$), where $q_c$,
is  the fastest growing mode,
\be
q_c=\l\sqrt{-\f{B+2\zeta_1\m}{D}}-1\r^{\f{1}{2}}\, .
\label{eq:qc}
\ee

The phase boundary between the {\it unstable oscillatory and contractile instability phases} can be obtained from the solution of $Im \[\lambda_{\pm}\]=0$,
\be
\zeta_1\m\Big\vert_{O \to CI} = \frac{1}{2} \left(-B + \sqrt{2 B k-k^2+2 B D q^2-2 D k q^2-D^2 q^4}\right)		\nn
\ee

Precisely at the stable-unstable phase boundary, since $Re \[\lambda_{\pm}\]=0$,
 the solutions correspond to left and right traveling waves for $\rho_b$ (and $u$), of the form 
 \be
\rho_b(x,t) = A \cos(q_c\,x \pm w_c\,t + \theta)
\label{eq:travelling_wave}
\ee
where the wavelength $q_c$ and frequency $w_c$ of the wave are given by, $q_c^* = \l\f{k}{D}\r^{\f{1}{4}}$ and $\omega^*_c = \sqrt{\f{k B}{2}}$,
leading to a wave-speed, $v_c^* \equiv \f{\omega_c^*}{q_c^*} = \l\f{kDB^2}{4}\r^{\f{1}{4}}$.

\subsection*{Strain dependent unbinding}
For the turnover of bound myosin filament density,  we allow for a possible strain-induced unbinding
 of the Hill-form,
$k_u = k_{u0} e^{\alpha \nabla \cdot {\bf u}}$. We find that the sign of coefficient $\alpha$ can be taken to be either positive or negative.
This affects the phase boundaries, but not the qualitative aspects of the phases. 
We find that the oscillatory phase exists, as long as  $\alpha<\alpha_{max}$, where
\bea
\alpha_{max} = \frac{BD + (B+D-ck)\zeta_1\m + (\zeta _1\m)^2}{k \zeta _1\m}
\eea


\begin {figure*}
\centering
\includegraphics[width=90mm]{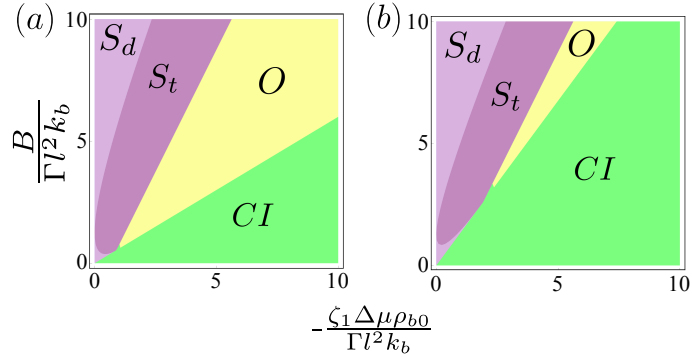}
\caption{Sign of $\alpha$ does not change the qualitative nature of the phase diagram in the effective elastic stress density versus active stress density
(nomalized to the frictional stress). The color scheme and legend are as in Fig.\,S1, (a) $\alpha=-0.5$ and (b) $\alpha=0.5$. Other parameters are $k=10$,
 $c=0.1$ and $D=0.1$.}
  \label{fig:alpha_sign}
\end{figure*}

\section*{Full nonlinear theory : additional material}

\begin {figure*}
\centering
\includegraphics[width=160mm]{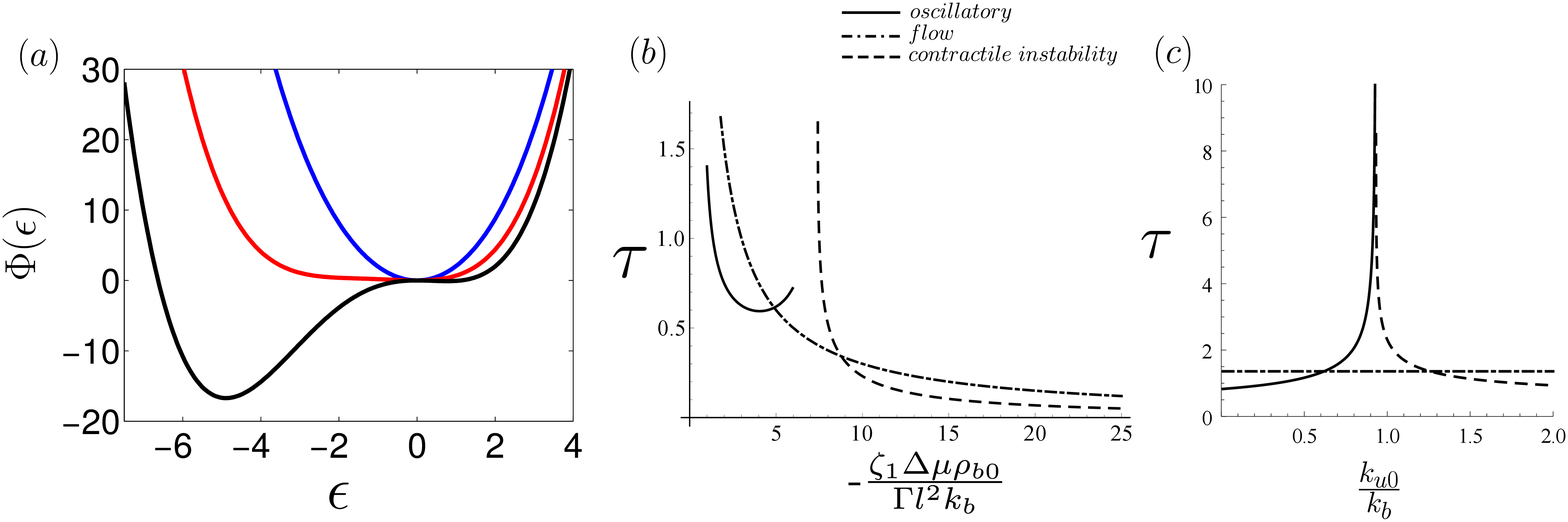}
\caption{ (a) Effective elastic free energy $\Phi(\epsilon)$ as a function of strain for three different values of active stress. At low active stress,
$\Phi$ has a single minimum at $\epsilon=0$, the free-energy profile has a lower curvature (corresponding to lower renormalized elastic modulus) than the passive elastomer.
At intermediate values of active stress, there appears a second minimum at $\epsilon=\epsilon_0$. At higher values of active stress, the minimum at $\epsilon=0$
becomes unstable.
(b) Time scale of various events - oscillation (dark line), traveling front (dot-dash line) and collapse (dash line) - as a function of the active stress, shows that with increasing active stress one first encounters the oscillatory phase,
then the traveling front and finally the collapse. The boundaries of these transitions support coexisting behaviours. (c) A similar time scale of events
 as a function of myosin inverse lifetime, $k_{u0}/k_b$. Legend same as in (b).
 }
	  \label{fig:understanding}
\end{figure*}


\begin {figure*}
\centering
\includegraphics[width=100mm]{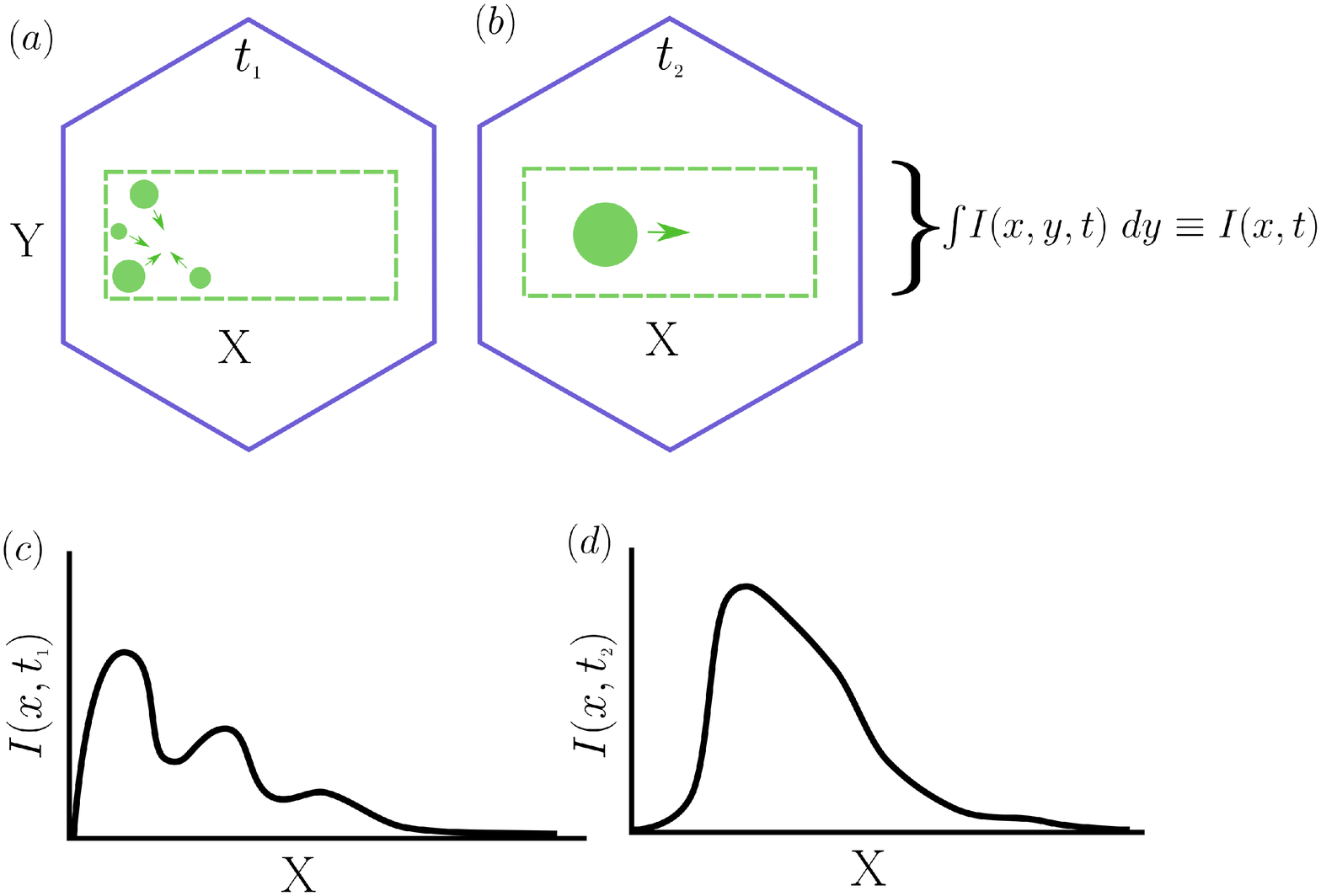}
\caption{Protocol for image analysis of the experiments on the dynamics of labeled medial myosin. (a,b) We collect intensity of myosin (green dots) at different time, e.g., $t_1, t_2$ (with $t_2>t_1$), within a thin rectangular strip (green rectangle)
chosen so that it is not contaminated by signals at the cell boundary.  Along each thin rectangular line (black), we integrate the intensity $I(x,y,t)$ along $y$, 
to obtain a 1-dimensional profile as shown.
(c,d) The $x$-profile of the one dimensional $I(x,t)$ is plotted at different times, $t_1$ and $t_2$. This example shows nucleation, coalescence and flow of the myosin-dense
regions. To obtain high contrast images, we background subtract the intensity maps.
	  }
	  \label{fig:image}
\end{figure*}

\begin {figure*}
\centering
\includegraphics[width=100mm]{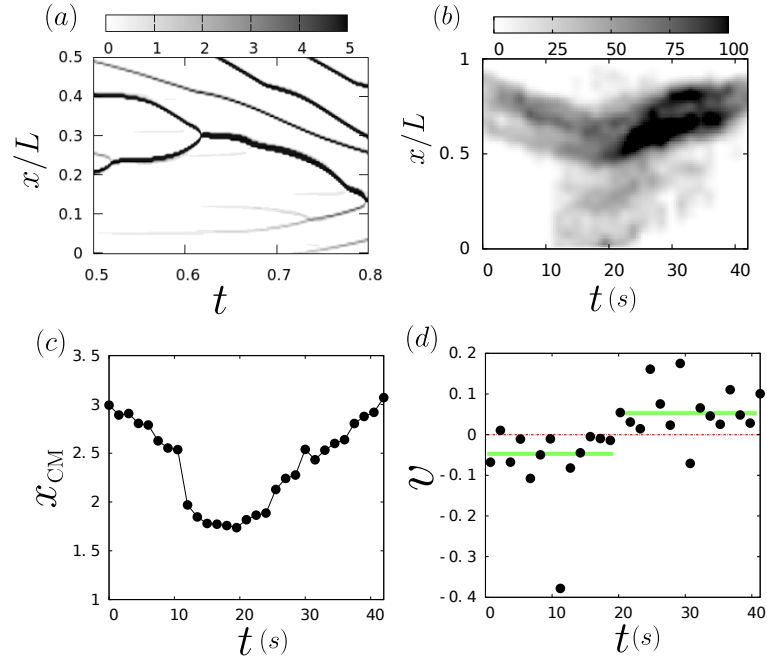}
\caption{Direction reversals of the traveling front indicate that the driving force for flow does not come from the boundary but  from the asymmetric profile of the
myosin-dense regions. (a) Direction reversals, seen in the kymographs obtained from numerical simulation of the full nonlinear dynamics, occur during certain
coalescence events. (b) Instance from the experimental myosin images where similar reversals are observed. This image is analyzed to show that (c) the movement of the centre of mass of the traveling front $x_{CM}$ changes direction upon coalescence, and (d) the velocity changes sign, but roughly has the same magnitude upon
coalescence.
}
	  \label{fig:reversals}
\end{figure*}

\clearpage

\section*{Supplementary Movies}

\noindent
{\bf Movie S1} : In-vivo imaging in a typical germ band cell of the drosophila wing imaginal disc, shows
nucleation, growth and coalescence of an actomyosin-dense cluster (GFP tagged myosin-RLC (WT)) and its subsequent flow towards a vertical cell junction.\\

\noindent
{\bf Movie S2}  : Numerically calculated dynamics of the spatial profile of bound myosin density (versus scaled coordinate $x/L$, where $L$ is the cell size) 
starting from statistically uniform initial conditions. Note the 
 nucleation, growth, coalescence of myosin enriched regions and subsequent formation of a spatially localized traveling front. 
Parameter values are $B=6$, $-\zeta_1\m=5.5$, $k=0.25$, $\alpha=1$, $D=0.25$. Periodic boundary conditions.\\

\noindent
{\bf Movie S3}  : Numerically calculated spatial profile of bound myosin density (red) and strain (blue) within the spatially localized traveling front.  Myosin contractility compresses the actin mesh locally, resulting in a localized extension region ahead of the leading edge of the traveling pulse. The traveling pulse maintains its shape as it moves.
Parameter values are $B=6$, $-\zeta_1\m=5.6$, $k=0.2$, $\alpha=1$, $D=0.25$. Periodic boundary conditions.


\begin{thebibliography}{9}

\footnotesize


\bibitem{Martin} A.C. Martin, M. Kaschube  and E.F. Wieschaus, Pulsed contractions of an actin-myosin network drive apical constriction, Nature
{\bf 457}, 495-499 (2009).
\bibitem{solonbrunner} J. Solon, A. Kaya-Copur, J. Colombelli and D. Brunner, Pulsed forces timed by a ratchet-like mechanism drive directed tissue movement during dorsal closure. Cell {\bf 137}, 1331-1342  (2009).
\bibitem{meghana} C. Meghana, N. Ramdas, F. M. Hameed, M. Rao, G. V. Shivashankar, and M. Narasimha,
 Integrin adhesion drives the emergent polarization of active cytoskeletal stresses to pattern cell delamination,
 Proc. Nat. Acad. Sc. {\bf 108}, 9107-9112 (2011).
 \bibitem{blanchard} G.B. Blanchard, S. Murugesu, R.J. Adams, A. Martinez-Arias, and N. Gorfinkiel,
 Cytoskeletal dynamics and supracellular organisation of cell shape fluctuations during dorsal closure,
Development {\bf 137}, 2743-2752 (2010).
 \bibitem{rohjohnson}M. Roh-Johnson, G. Shemer, C.D. Higgins, J.H. McClellan, A.D. Werts, U.S. Tulu, L. Gao, E. Betzig, D.P. Kiehart, and B. Goldstein,
 Triggering a cell shape change by exploiting preexisting actomyosin contractions, Science {\bf 335}, 1232-1235 (2012).
 
 \bibitem{Lecuit2010}
M. Rauzi, P.F. Lenne, and T. Lecuit, Planar polarized actomyosin contractile flows control epithelial junction remodelling,
Nature {\bf 468}, 1110-1114 (2010). 
 \bibitem{Lecuit2013}R. Levayer and T. Lecuit, Oscillation and Polarity of E-Cadherin Asymmetries Control Actomyosin Flow Patterns during Morphogenesis,
Dev. Cell {\bf 26}, 162-175 (2013).

\bibitem{akankshi}A. Munjal, J-M. Philippe, E. Munro	 and T. Lecuit, 
A self-organized biomechanical network drives shape changes during tissue morphogenesis, Nature {\bf 524}, 351-355 (2015).

\bibitem{zallen}R. Fernandez-Gonzalez, and J.A. Zallen,
Oscillatory behaviors and hierarchical assembly of contractile structures in intercalating cells,
Phys Biol. {\bf 8}, 045005 (2011).


\bibitem{Lecuit2008} M. Rauzi, P. Verant, T. Lecuit, and P.F. Lenne,
Nature and anisotropy of cortical forces orienting Drosophila tissue morphogenesis,
Nat. Cell Biol. {\bf 10}, 1401-1410 (2008).

\bibitem{xenopus} R. Keller, L. Davidson, A. Edlund, T. Elul, M. Ezin, D. Shook and P. Skoglund,
 Mechanisms of convergence and extension by cell intercalation, Phil. Trans. R. Soc. Lond. {\bf 355}, 897-922 (2000).
 

\bibitem{Martin_jcb2010} A. C. Martin, M. Gelbart, R. Fernandez-Gonzalez, M. Kaschube and E. F. Wieschaus1,
Integration of contractile forces during tissue invagination,
J. Cell Biol. {\bf 188}, 735-749 (2010).



\bibitem{Wu_NCB_2014} S. K. Wu, G. A. Gomez, M. Michael, S. Verma, H. L. Cox, J. G. Lefevre, R. G. Parton, N. A. Hamilton, Z. Neufeld and A.S. Yap,
Cortical F-actin stabilization generates apical–lateral patterns of junctional contractility that integrate cells into epithelia,
Nat. Cell Biol. {\bf 16}, 167–178 (2014)

\bibitem{cavey}M. Cavey, M. Rauzi, P.F. Lenne, and T. Lecuit,
A two-tiered mechanism for stabilization and immobilization of E-cadherin.
Nature {\bf 453}, 751-756 (2008).

\bibitem{yonemura}S. Yonemura, Y. Wada, T. Watanabe, A. Nagafuchi, and M. Shibata,
alpha-Catenin as a tension transducer that induces adherens junction development,
Nat. Cell Biol. {\bf 12}, 533-542 (2010).

\bibitem{buckley}C.D. Buckley, J. Tan, K.L. Anderson, D. Hanein, N. Volkmann, W.I. Weis, W.J. Nelson, and A.R. Dunn,
The minimal cadherin-catenin complex binds to actin filaments under force,
Science {\bf 346}, 1254211 (2014).






%

\bibitem{Marchetti2011a}  S. Banerjee, T.B. Liverpool and M.C. Marchetti,
Generic phases of cross-linked active gels: Relaxation, oscillation and contractility, 
Europhys. Lett. {\bf 96}, 58004 (2011).


\bibitem{kovacs} M. Kovacs, K. Thirumurugan, P. J. Knight and J. R. Sellers,
Load-dependent mechanism of nonmuscle myosin 2, 
Proc. Nat. Acad. Sc. {\bf 104}, 9994 –9999 (2007)

\bibitem{Eaton_devCell} R. Fernandez-Gonzalez, S. de M. Simoes, J.C. Röper, S. Eaton and J. A. Zallen,
Myosin II dynamics are regulated by tension in intercalating cells,
Dev. Cell {\bf 17}, 736–743 (2009).

\bibitem{Marchetti2011b} S. Banerjee and M.C. Marchetti, Instabilities and oscillations in isotropic active gels.
Soft Matter {\bf 7}, 463 (2011).

\bibitem{keener}J. Keener and J. Sneyd, in {\it Mathematical Physiology, I: Cellular Physiology}, (2nd Ed., Springer, NY, 2009).

\bibitem{SolonSalbreux} K. Dierkes, A. Sumi, J. Solon and G. Salbreux, Spontaneous oscillations of 
elastic contractile materials with turnover, Phys. Rev. Lett.  {\bf 113}, 148102 (2014).


\bibitem{rmp}M. C. Marchetti, J.-F. Joanny, S. Ramaswamy, T. B. Liverpool, J.
 Prost, M. Rao and R.A. Simha, Hydrodynamics of Soft Active Matter, Rev.\,Mod.\,Phys. {\bf 85} 1143 (2013).
 
\bibitem{Vasquez_2014} C. G. Vasquez, M. Tworoger and A. C. Martin,
Dynamic myosin phosphorylation regulates contractile pulses and tissue integrity during epithelial morphogenesis,
J. Cell Biol. {\bf 206}, 435-450 (2014). 
 
\bibitem{Kasza_KE_PNAS14} K. E. Kasza, D. L. Farrell and J. A. Zallen,
Spatiotemporal control of epithelial remodeling by regulated myosin phosphorylation,
Proc. Nat. Acad. Sc. {\bf 111}, 11732-11737 (2014).
 
\bibitem{lenne_PNAS15} K. Bambardekar, R. Clement, O. Blanc, C. Chardes, and P.F. Lenne,
Direct laser manipulation reveals the mechanics of cell contacts in vivo,
Proc. Nat. Acad. Sc. {\bf 112}, 1416-1421 (2015).

\bibitem{strogatz}S. Strogatz in {\it Nonlinear dynamics and Chaos}, (2nd Ed., Perseus, Cambridge, MA (2014)).

\bibitem{debsankar}D. Banerjee et al., in preparation.


\bibitem{Mayer_Nature_2010} M. Mayer, M. Depken, J.S. Bois, F. Julicher and  S.W. Grill, Anisotropies in cortical
tension reveal the physical basis of polarizing cortical flows,
Nature {\bf 467}, 617-621 (2010).

\bibitem{muthukumar}A.K. Murthy and M. Muthukumar, Dynamic mechanical properties of poly ($\gamma$-benzyl L-$\alpha$-glutamate) gels in benzyl alcohol, 
Macromolecules {\bf 20}, 564-569 (1987).

\bibitem{leibler} L. Leibler, M. Rubinstein and R.H. Colby, Dynamics of reversible networks, Macromolecules {\bf 24}, 4701-4707 (1991).

\bibitem{tanakaedwards}F. Tanaka and S.F. Edwards, Viscoelastic properties of physically crosslinked networks: Transient network theory,
Macromolecules {\bf 25}, 1516-1523 (1992); Viscoelastic properties of physically crosslinked networks: Non-linear stationary viscoelasticity,
J. Non-Newtonian Fluid Mech. {\bf 43} 247-271 (1992).

\bibitem{munro}T. Kim, M.L. Gardel and E. Munro, Determinants of Fluidlike Behavior and Effective Viscosity in Cross-Linked
Actin Networks, Biophys J. {\bf 106}, 526-534 (2014).

\bibitem{grill1}N. W. Goehring, P. Khuc Trong, J. S. Bois, D. Chowdhury, E. M. Nicola, A. A. Hyman, and S. W. Grill, Polarizing PAR proteins by advective triggering of a pattern-forming system, Science {\bf 334}, 1137-1141 (2011). 














\bibitem{kamm}M. Mak, M.H. Zaman, R.D. Kamm and T. Kim, Interplay of active processes
modulates tension and drives phase transition in self-renewing, motor-driven cytoskeletal networks,
Nat. Comm. {\bf 7}, 10323 (2016).

\bibitem{murrell}W. Jung, M.P. Murrell and T. Kim,
F-actin cross-linking enhances the stability of force generation in disordered actomyosin networks,
Comp. Part. Mech. {\bf 2}, 317-327 (2015).


\bibitem{Sweby1984} P. K. Sweby, High Resolution Schemes Using Flux Limiters for Hyperbolic Conservation Laws,
SIAM J. Num. Anal. {\bf 21}, 995-1011 (1984).

\bibitem{leveque}R.J. LeVeque in {\it Numerical Methods for Conservation Laws}, Lectures in Mathematics (2nd edition, Berlin, Birkhauser 1992).

\bibitem{Courant1952} R. Courant, E. Isaacson and M. Rees, On the solution of nonlinear hyperbolic differential equations by finite differences,
Comm. Pure App. Math. {\bf V}, 243-255 (1952).

\bibitem{Lax1960} P. Lax and B. Wendroff, Systems of conservation laws,
Comm. Pure App. Math. {\bf XIII}, 217-237 (1960).

\bibitem{tvdRK} S. Gottlieb and C-W. Shu, Total variation diminishing Runge-Kutta scheme,
Math. Comp. {\bf 67}, 73-85 (1988).

\bibitem{Cavey&Lecuit} M. Cavey, and T. Lecuit,  Imaging cellular and molecular dynamics in live embryos using fluorescent proteins,
Meth. Mol Biol {\bf 420}, 219-238 (2008).


\end{thebibliography}

\begin{thebibliography}{9}

\footnotesize

\bibitem{Marchetti2011a}  S. Banerjee, T.B. Liverpool and M.C. Marchetti,
Generic phases of cross-linked active gels: Relaxation, oscillation and contractility, 
Europhys. Lett. {\bf 96}, 58004 (2011).


\bibitem{Marchetti2011b} S. Banerjee and M.C. Marchetti, Instabilities and oscillations in isotropic active gels,
Soft Matter {\bf 7}, 463 (2011).

\bibitem{akankshi}A. Munjal, J-M. Philippe, E. Munro	 and T. Lecuit, 
A self-organized biomechanical network drives shape changes during tissue morphogenesis, Nature (2015).


\bibitem{Lecuit2010}
M. Rauzi, P-F. Lenne and T. Lecuit, Planar polarized actomyosin contractile flows control epithelial junction remodelling,
Nature {\bf 468}, 1110-1114 (2010). 

\bibitem{SolonSalbreux} K. Dierkes, A. Sumi, J. Solon and G. Salbreux, Spontaneous oscillations of 
elastic contractile materials with turnover, Phys. Rev. Lett.  {\bf 113}, 148102 (2014).

\bibitem{Mayer_Nature_2010} M. Mayer, M. Depken, J.S. Bois, F. Julicher and S.W. Grill, Anisotropies in cortical
tension reveal the physical basis of polarizing cortical flows,
Nature {\bf 467}, 617-621 (2010).


\bibitem{solonbrunner} J. Solon, A. Kaya-Copur, J. Colombelli and D. Brunner, Pulsed forces timed by a ratchet-like mechanism drive directed tissue movement during dorsal closure,
Cell {\bf 137}, 1331-1342  (2009).


\bibitem{Sweby1984} P. K. Sweby, High Resolution Schemes Using Flux Limiters for Hyperbolic Conservation Laws,
SIAM Journal on Numerical Analysis {\bf 21}, 995-1011 (1984).

\bibitem{leveque}R.J. LeVeque in {\it Numerical Methods for Conservation Laws}, Lectures in Mathematics (2nd edition, Berlin, Birkhauser 1992).

\bibitem{Courant1952} R. Courant, E. Isaacson and M. Rees, On the solution of nonlinear hyperbolic differential equations by finite differences,
Comm. Pure App. Math. {\bf V}, 243-255 (1952).

\bibitem{Lax1960} P. Lax and B. Wendroff, Systems of conservation laws,
Comm. Pure App. Math. {\bf XIII}, 217-237 (1960).

\bibitem{tvdRK} S. Gottlieb and C-W. Shu, Total variation diminishing Runge-Kutta scheme,
Math. Comp. {\bf 67}, 73-85 (1988).



\end{thebibliography}
\end{document}